%% file: main_v2.tex
\documentclass[10pt]{IEEEtran_v15}
\usepackage{psfig,amssymb,amsmath,header,color}
\begin{document}

\title{Correlated Link Shadow Fading in Multi-hop Wireless Networks}
\author{Piyush Agrawal and Dr. Neal Patwari\\
    Sensing and Processing Across Networks Lab\\
    Department of Electrical and Computer Engineering\\
    University of Utah, Salt Lake City, USA\\
    \texttt{[pagrawal, npatwari]@ece.utah.edu}\\
}
\maketitle

\begin{abstract}

Accurate representation of the physical layer is required for
analysis and simulation of multi-hop networking in sensor, ad hoc,
and mesh networks.  This paper investigates, models, and analyzes
the correlations that exist in shadow fading between links in
multi-hop networks. Radio links that are geographically proximate
often experience similar environmental shadowing effects and thus
have correlated fading.  We describe a measurement procedure and
campaign to measure a large number of multi-hop networks in an
ensemble of environments. The measurements show statistically
significant correlations among shadowing experienced on different
links in the network, with correlation coefficients up to 0.33. We
propose a statistical model for the shadowing correlation between
link pairs which shows strong agreement with the measurements, and
we compare the new model with an existing shadowing correlation
model of Gudmundson (1991). Finally, we analyze multi-hop paths in
three and four node networks using both correlated and independent
shadowing models and show that independent shadowing models can
underestimate the probability of route failure by a factor of two or
greater.
\end{abstract}








\section{Introduction}
\label{S:Introduction}
\input{intro_model_v2}


\section{Measurement Setup}
\label{S:MeasurementSetup}
\input{NCMS_description}

\section{Experiment}
\label{S:Experiment}
\input{NCMS_experiment}

\section{Statistical Analysis}
\label{S:StatisticalAnalysis}
\input{StatisticalAnalysis}

\section{Joint Path Loss Model}
\label{S:CorrelationModelDes}
\input{model_corr_derivation}

\section{Application of Joint Model}
\label{S:application_model}
\input{application_model}

\section{Conclusion}
\label{S:conclusion}
\input{conclusion}

\section*{APPENDIX}
\label{S:appendix}
\input{appendix}

\bibliography{allRef}
\bibliographystyle{IEEEtran}
\newpage

\begin{figure}[htbp]
  \centerline{
  \psfig{figure=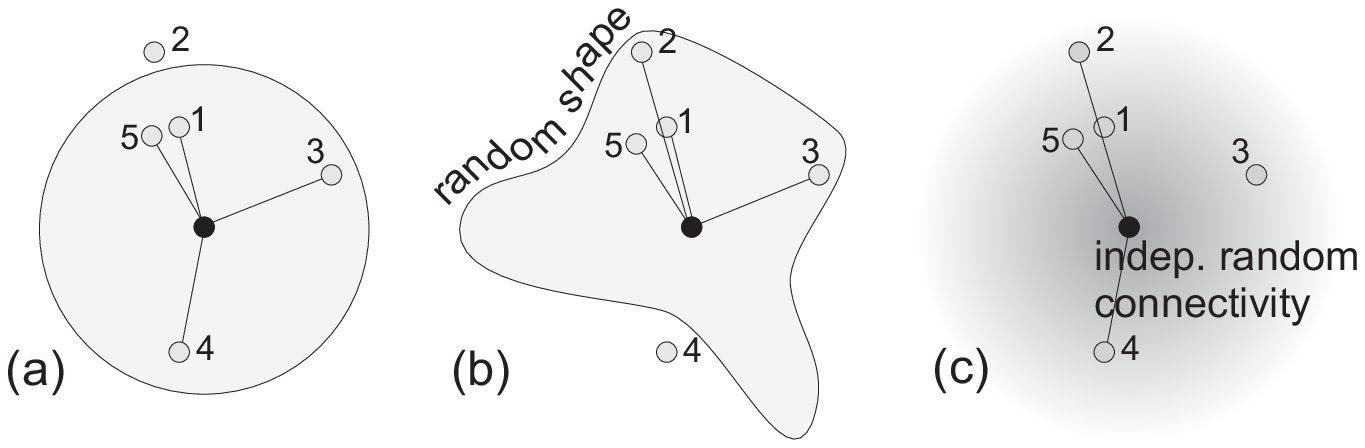,width= 3.5in}}
  \caption{Graphical depiction of (a) circular coverage model, and (c) coverage in the i.i.d.~log-normal
  shadowing model, compared to the common depiction of (b) in which coverage area is a random shape.
  In (a) and (b), nodes are connected if and only if they are within the gray area, while in (c), nodes are connected
  with probability proportional to the shade (darker is more probable).}
  \label{F:coverageModel}
\end{figure}

\begin{figure}[htbp]
  \centerline{
  \psfig{figure=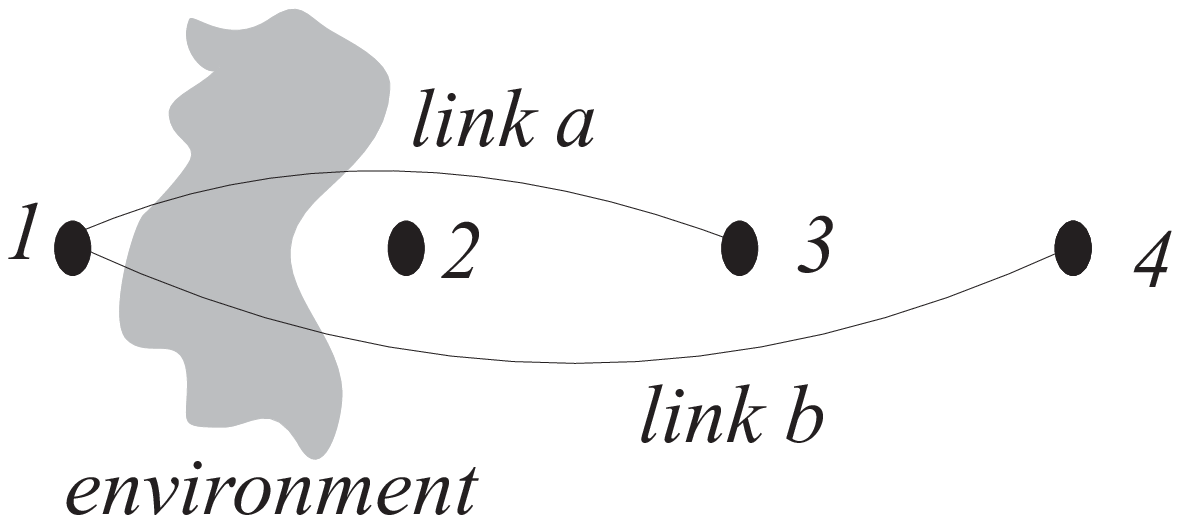,width=2in}}
  \caption{Example of factor in shadowing loss correlation.
  Because link $a$ and link $b$ cross the same
  environment, their shadowing losses tend to be correlated.}
  \label{F:Examples}
\end{figure}

\begin{figure}[htbp]
\centerline{
    \psfig{figure=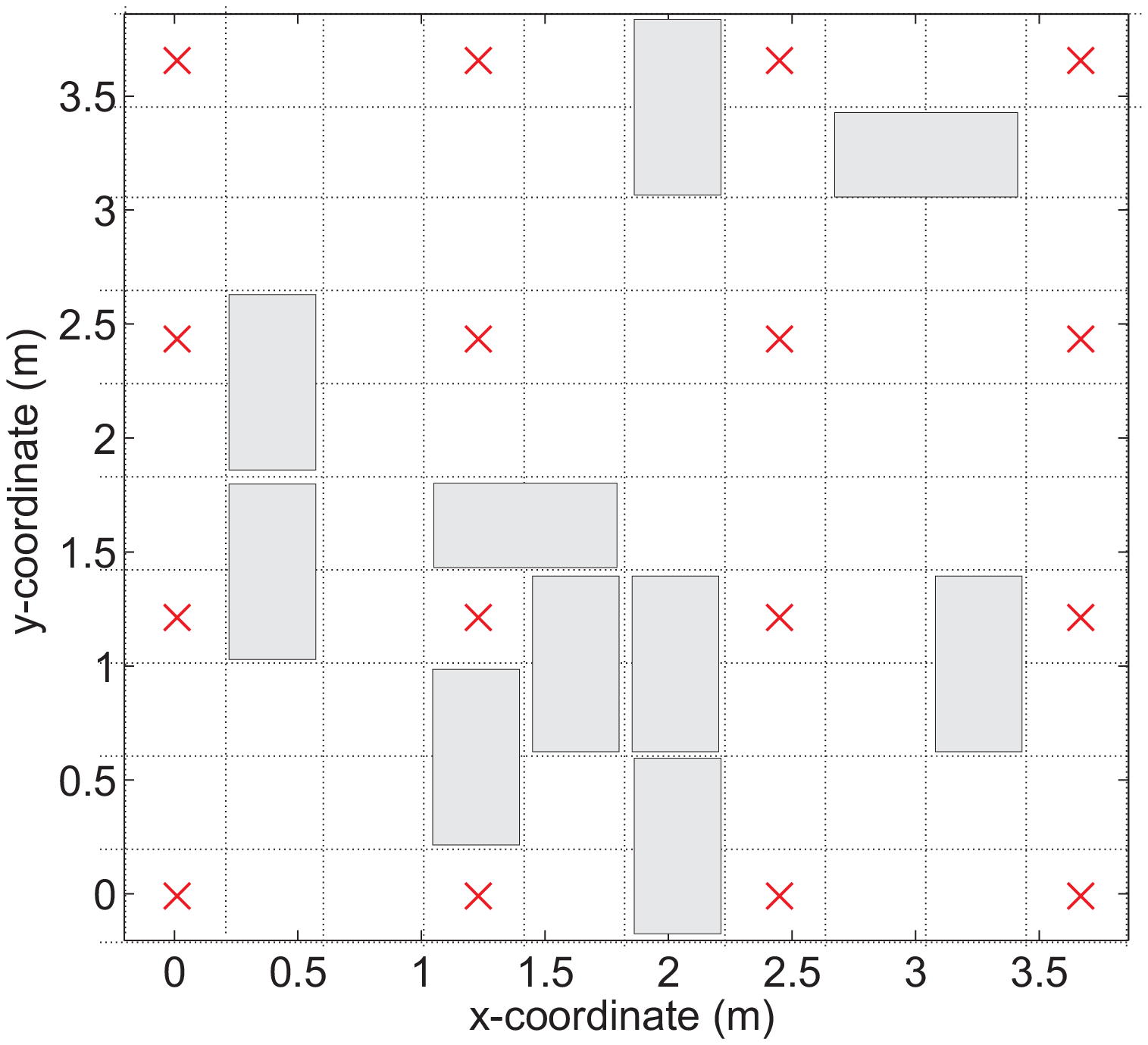,width=2.4in}
    }
\caption{One realization of the random locations of the boxes among
the 16 node locations (\textcolor{red}{$\times$}). Each box (grey rectangles) occupies two pixels of this
graph and can be placed either parallel or perpendicular to x-axis.}
\label{F:possibleBoxPlacements}
\end{figure}

\begin{figure}[htbp]
  \centerline{
  \psfig{figure=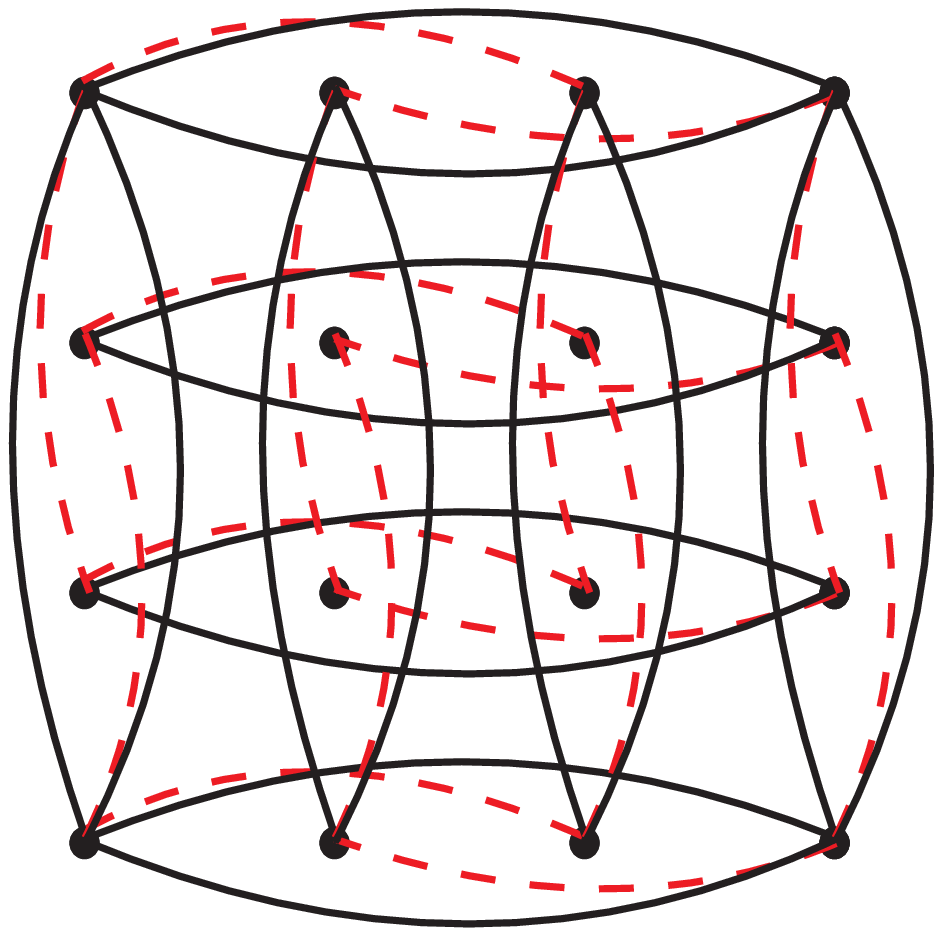,width=1.4in}}
  \caption{Link pairs with identical link geometry in a grid deployment. Each link pair
is shown with one link as a dotted (\textcolor{red}{-~-}) line and another link as a
solid lines (--). All link pairs with identical link geometry are
shown.}
  \label{F:SimilarGeoLink}
\end{figure}

\begin{figure}[htbp]
  \centerline{
  \psfig{figure=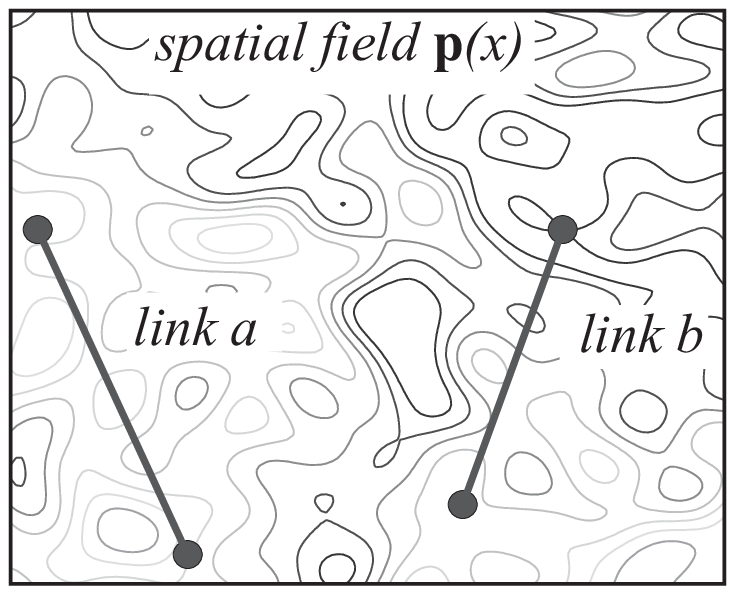,width=3in}}
  \caption{A link pair in an underlying spatial loss field}
  \label{F:linkPairInRandomField}
\end{figure}

\begin{figure}[htbp]
  \centerline{
  \psfig{figure=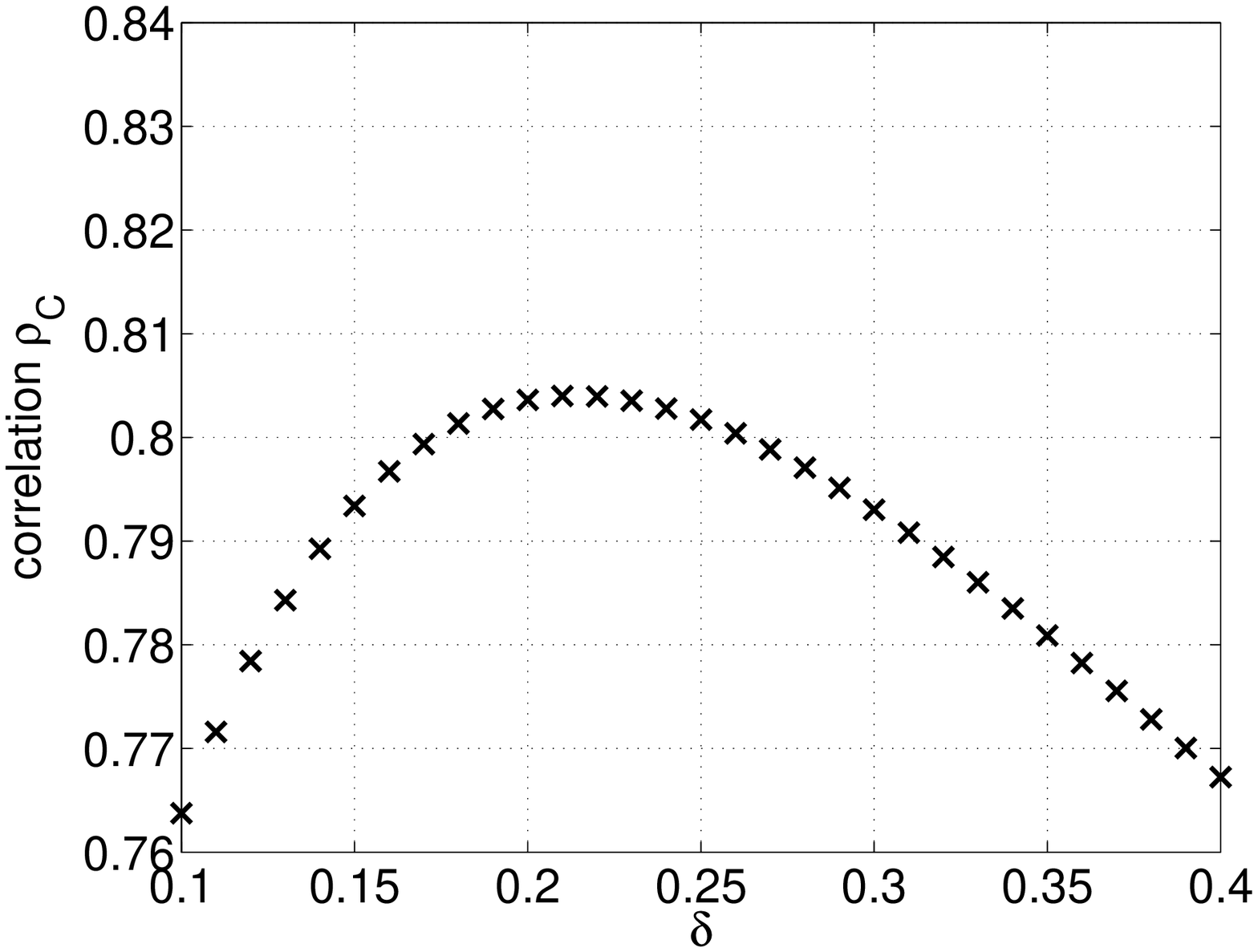,width=2.7in}}
  \caption{Variation of $\rho_C$ with $\delta$. }
  \label{F:corr_varyingDelta}
\end{figure}

\begin{figure}[htbp]
  \centerline{
  \psfig{figure=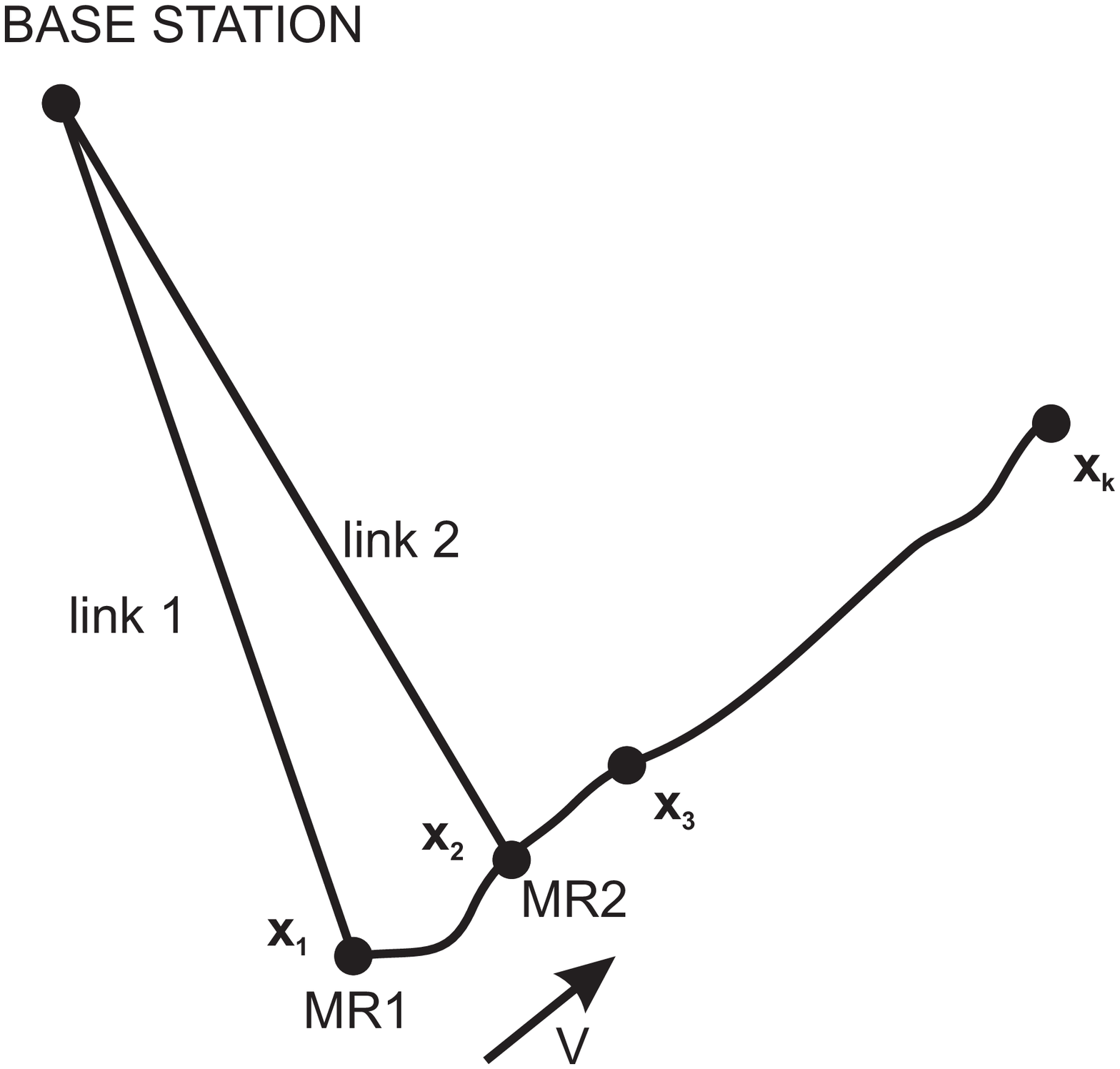,width=2.5in}}
  \caption{Example of the motion of mobile receiver and base station
  position in Gudmundson's model}
  \label{F:mobileBSPosition}
\end{figure}

\begin{figure}[htbp]
\centerline{(a)  \psfig{figure=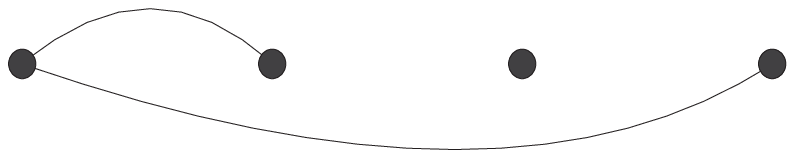,width=1.5in} $\quad$
(b)  \psfig{figure=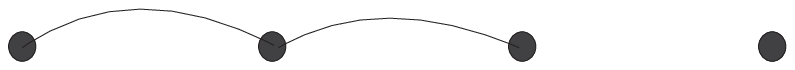,width=1.5in}
}
  \caption{A case of two different types of links, shown by (a)
  and (b). The Gudmundson's model predicts identical correlation for the two
  cases while the proposed model does not. Experimentally, the
  correlations vary significantly from (a) 0.21 to (b) 0.05}
  \label{F:limit_Gudmundson}
\end{figure}

\begin{table}[htbp]
\begin{center}
\begin{tabular}{|l|c|}
  \hline
  & Correlation with Measured Data\\
  \hline
  Proposed Model & 0.804 \\
  \hline
  Gudmundson's Model & 0.644 \\
  \hline
\end{tabular}
\caption{Comparison between the proposed model and Gudmundson's
model}\label{T:comapr_propModelNGudmundsonModel}
\end{center}
\end{table}

\begin{figure}[htbp]
\centerline{
(a)  \psfig{figure=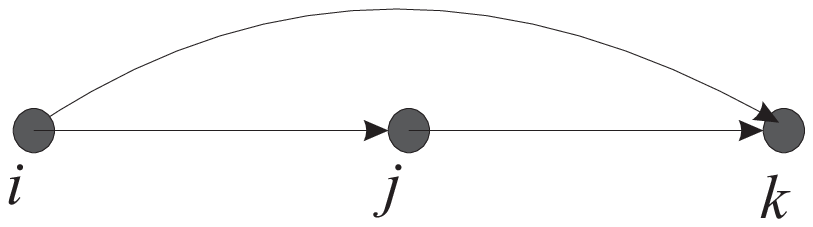,width=1.5in} $\quad$
(b)    \psfig{figure=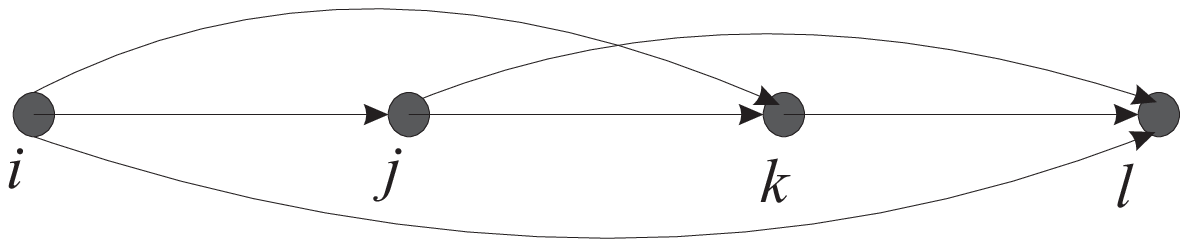,width=2.2in}}
  \caption{Example multi-hop networks of (a) three nodes and (b) four nodes.}\label{F:simpleLink}
\end{figure}

\begin{figure}[htbp]
\centerline{
  \psfig{figure=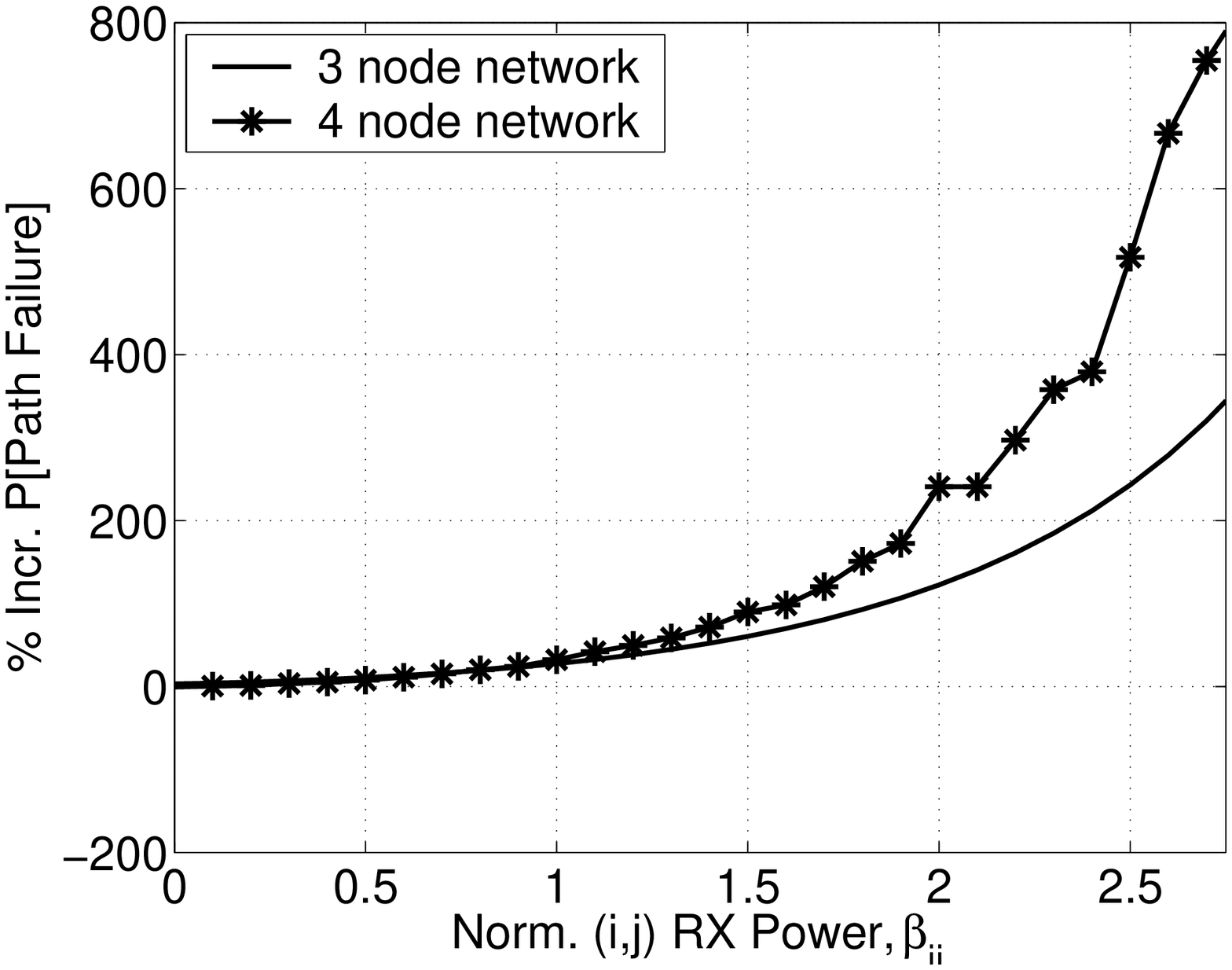,width=3in}}
  \caption{Plot showing the variation of percentage increment in the $\PR{\text{link failure}}$ for a
  3 node and 4 node multi-hop network with normalized received power of the shortest link $(i,j)$, $\beta_{i,j}$.}
  \label{F:PercentageIncrProb_3n4}
\end{figure}

\end{document}

%% file: intro_model_v2.tex

Both simulation and analysis are critical to the development of
multi-hop networks, including mesh, ad hoc, and sensor networks.
However, current physical layer models do not accurately represent
radio channels in multi-hop wireless networks \cite{lee07}, and as a
result, there is a significant disconnect between simulation and
real world deployment. There is significant interest in improving
statistical models beyond the current state-of-the-art in order to
decrease the difference between simulation and analysis results and
experimental deployment results.

This paper presents a statistical joint path loss model between a
set of static nodes. Joint path losses and transmit powers determine
the connectivity, reliability in interference, and energy
consumption during power control, of network communications. Channel
models used in multi-hop networks have considered path losses to be
independent, yet they are correlated through shadowing effects.  We
demonstrate these correlations via measurements and present a
correlated shadowing loss model, which is then shown to have a
dramatic effect on network connectivity.

We do not address other random processes like transmit power
variation, manufacturing variations among nodes, position of nodes
in random deployments, mobility of nodes, or interference models.
However, the developed model informs future development of path loss
models for mobile networks, and may be used to analyze the effects
of other variation and interference models.

\subsection{Single-Link Path Loss Model}

Radio propagation measurement and modeling for a single radio link
has been reported extensively over the past century
\cite{marconi1899,hashemi93,rappaport96,bertoni1999rpm}. In general,
when there is no \emph{site-specific} knowledge of the environment,
the ensemble mean received power, $\bar{P}(d)$ (dBm), at a distance
$d$ from the transmitter, is \cite{hashemi93,rappaport96},
\begin{equation} \label{E:meanPathLoss}
 \bar{P}(d) = P_{T} - \Pi_0 - 10 n_p \log_{10} \frac{d}{\Delta_0},
\end{equation}
where $P_{T}$ is the transmitted power in dBm, $n_p$ is the path
loss exponent, and $\Pi_0$ is the loss experienced at a short
reference distance $\Delta_0$ from the transmitter antenna. This
model incorporates the free space path loss model when $n_p = 2$,
and extends to practical (obstructed) multipath environments when
$n_p > 2$.

On a particular link, received power will vary from the ensemble
mean because of \emph{fading}. The measured received power for the
link between transmitter $i$ and receiver $j$ is,
\begin{equation} \label{E:logNormalShadowing}
 P_{i,j} = \bar{P}(d_{i,j}) - Z_{i,j},
\end{equation}
where $d_{i,j}$ is the distance between nodes $i$ and $j$, and
$Z_{i,j}$ is the fading loss. In general, shadow fading, small-scale
or frequency-selective fading, and antenna and device losses all
contribute to $Z_{i,j}$. Wideband receivers reduce the effects of
small-scale or frequency-selective fading issues, and antenna and
device-caused variations are generally small compared to shadowing
variations. Shadow fading, also called medium-scale fading
\cite{hashemi93}, describes the loss suffered as the signal passes
through or diffracts around major obstructions in its path from the
transmitter to the receiver. These obstructions include walls and
furniture indoors, and buildings, terrain, and trees outdoors.

We hypothesize that shadowing losses are correlated across different
links which are geographically proximate. Since shadowing is central
to the analysis in this paper, we separate total fading loss
$Z_{i,j}$ into two contributions,
\begin{equation}\label{E:shadowFadingDefn}
    Z_{i,j} = X_{i,j} + Y_{i,j},
\end{equation}
where $X_{i,j}$ represents the shadowing loss, and $Y_{i,j}$
represents all other (non-shadowing) losses.

\subsection{Application in Multi-hop Networking Research}

In the multi-hop networking simulation and analysis literature, two
models are used:
\begin{enumerate}
  \item The circular coverage model: $Z_{i,j}=0$ for all links, and thus
  the coverage area is a perfect circle, as shown in Figure
  \ref{F:coverageModel}(a).
  \item The i.i.d.~log-normal shadowing model: For all links $(i,j)$, random variables $Z_{i,j}$
  are independent and identically distributed Gaussian with zero mean and variance
  $\sigma_Z^2$, as shown in Figure
  \ref{F:coverageModel}(c).
\end{enumerate}
\showfigure{
\begin{figure}[htbp]
  \centerline{
  \psfig{figure=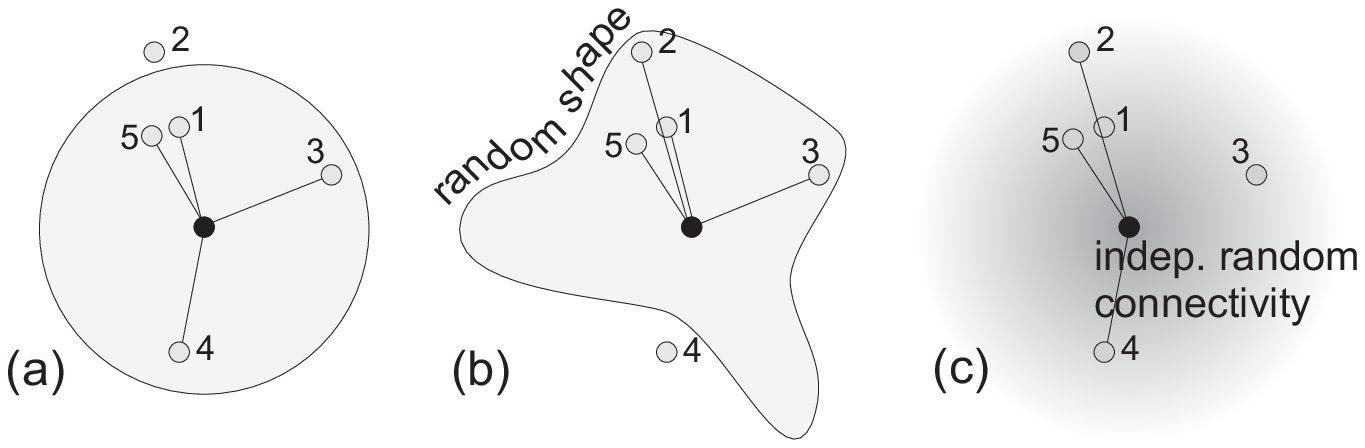,width= 4 in}}
  \caption{Graphical depiction of (a) circular coverage model, and (c) coverage in the i.i.d.~log-normal
  shadowing model, compared to the common depiction of (b) in which coverage area is a random shape.
  In (a) and (b), nodes are connected if and only if they are within the gray area, while in (c), nodes are connected
  with probability proportional to the shade (darker is more probable).}
  \label{F:coverageModel}
\end{figure}
}

We argue that both models are at opposite extremes, and both
problematic.  Note that `realistic coverage' is commonly depicted
pictorially as a coverage area with random range as a function of
angle \cite{bettstetter05,hekmat06}, as in Figure
\ref{F:coverageModel}(b), and neither fading model produces such a
random shape.  It is easy to recognize that the deterministic,
circular coverage areas are unrealistic for wireless communications
links. However, circular coverage has been a common assumption in ad
hoc and sensor network research and has been used to generate
foundational research results.  Kotz, Newport, and Elliot
\cite{kotz03} examined the set of papers in the MobiCom proceedings
from 1995 through 2002, and found that out of 36 papers which
required radio models, only four did not use a circular coverage
model.

In comparison, the i.i.d.~shadowing model is non-deterministic, and
eliminates the concept of coverage area.  Since the model has no
spatial memory, even two nearly overlapping links would be
represented as statistically independent.  For example node 2 in
Figure \ref{F:coverageModel}(c) may be connected while node 1 is
not.

Recent research, including Hekmat and Van Mieghem \cite{hekmat06}
and Bettstetter and Hartmann \cite{bettstetter05}, has studied
connectivity in ad hoc networks using the i.i.d.~log-normal
shadowing model. Their analyses indicate that for a constant level
of connectivity, node deployment density can be reduced when the
variance of the shadowing is increased. This increase in
connectivity is largely a result of the model's independence
assumption.  Since losses in links in the same direction from a
transmitter are independent, if one link is disconnected because of
high loss, another node in the same direction is likely to be
connected.

In reality, if an obstacle in one direction from a transmitter
strongly attenuates its signal, any receiver behind the obstacle is
likely to experience high fading loss. For example, if the
environment in Figure \ref{F:Examples} causes severe shadowing, it
is likely to cause additional path loss on both links $a$ and $b$.
In contrast, the i.i.d.~log-normal shadowing model assumes that the
shadowing across links $a$ and $b$ will be independent and thus
exaggerate the connectivity. We quantify this argument in Section
\ref{S:application_model}.

\subsection{Correlation Limits Link Diversity}
Diversity methods are common means to achieve reliability in
unreliable channels. Multi-hop networking serves as a network-layer
diversity scheme by allowing two nodes to be connected by any one of
several multi-hop paths. All diversity schemes are limited by
channel correlations.  Correlations have been studied and shown to
limit diversity gains in time, space, frequency and multipath
diversity schemes \cite{Durgin,hashemi93,rappaport96,saleh87}.

Yet little research has addressed channel correlations on links in
sensor, mesh, and ad hoc networks.  This paper presents an initial
investigation into quantitatively assessing the correlation in the
shadow fading experienced on the different links of a multi-hop
network.  This investigation is experimental, using full link
measurements of an ensemble of deployed networks to estimate and
test for statistical correlations. We propose a joint path loss
model which accurately represents observed correlations in link
shadowing. Further, we quantify the effect that such correlation has
on source to destination path statistics. We show that for a simple
three node network that the probability of path failure is double
what would be predicted by the i.i.d.~log-normal shadowing model.

\section{Related Work} \label{S:RelatedWork}

Shadow fading correlations have been measured and shown to be
significant in other wireless networks. For example: (1.) in digital
broadcasting, links between multiple broadcast antennas to a single
receiver have correlated shadowing which affects the coverage area
and interference characteristics \cite{malmgren97}; (2.) in indoor
WLANs correlated shadowing is significant (as high as 0.95) strongly
impacts system performance \cite{butterworth00}; and (3.) in
cellular radio correlation on links between a mobile station and
multiple base stations significantly affects mobile hand-off
probabilities and co-channel interference ratios
\cite{klingenbrunn99,zhang01,safak91}.

In cellular radio, the model of Gudmundson \cite{Gudmundson91} is
used to predict shadowing correlation for the link between a mobile
station (MS) to a base station over time as the MS moves. In
Section~\ref{S:CorrelationModelDes}, we address the difficulty in
applying this model to multi-hop networks.  We quantitatively
compare it with the proposed model when the Gudmundson model may be
applied.  Wang, Tameh, and Nix \cite{wang06} extended Gudmundson's
model to the case of simultaneous mobility of both ends of the link,
for use in MANETs,  and relate a sun-of-sinusoids method to generate
realizations of the shadowing process in simulation.  Both works use
``correlated shadowing'' to refer to the correlation of path loss in
a \emph{single link} over time, while the present work studies the
correlation of \emph{many disparate links} at a single time.

The closest study to the present work used RSS measurements in a
single network to quantify correlations between two links with a
common node \cite{M2MRadioPatwari}. Those results could not be
complete because a single measured network cannot provide
information about an ensemble of network deployments. The present
study uses multiple measured networks to examine many pairs of links
with the identical geometry, both with and without a common node.

%% file: NCMS_description.tex

In this section we present our method for measuring the path-loss of
each pair-wise link in a multi-hop network, using a specialized
sensor network. This system is referred to as the \textit{network
channel measurement system} (NCMS).  The NMCS allows us to quickly
measure the received power $P_{i,j}$ (dBm) of every link $(i,j)$ in
a deployed network, to measure across a range of frequencies, and to
record the data on a laptop for later analysis.


\subsection{Equipment}
The nodes used in the measurement campaign are ``mica2'' motes
manufactured by Crossbow. A mica2 mote operates in the 902 - 928 MHz
band using a Chipcon CC1000 FSK transceiver. The transmit power is
user programable and can be varied based on the network topology and
environmental density. The mica2 measures and reports RSS values for
each received signal \cite{micadatasheet}.


\subsubsection{Battery Variations}\label{S:battVariations}
Transmit power is proportional to the battery voltage squared. Since
measurement and data collection of one deployed network take on the
order of minutes, battery voltage can be considered constant during
each experiment. Each device measures and reports its own battery
voltage, and we monitor to ensure that battery voltages are largely
the same across devices throughout the experiment.



\subsection{Protocol}\label{S:protocolDescription}


\subsubsection{Software}
A NesC/TinyOS embedded program is written to operate the following
protocol:

\paragraph*{Frequency Hopping}
From the 902-928 MHz band, 14 center frequencies are chosen. The
nodes are programmed such that each node hops across all the 14
frequencies in each cycle. The time duration between frequency hops
is three seconds.

\paragraph*{Synchronization}
Synchronization is required so that frequency-hopping sensors are
all transmitting and receiving on the same frequency at the same
time.   One of the frequencies in the frequency band is considered
as a synchronization frequency and is repeated three times each
cycle so that neighbors can synchronize with each other more
quickly. The dwell time on each frequency includes a period in which
all sensors transmit a packet and receive packets from other
sensors, and a period for switching frequency.

\paragraph*{Pairwise measurements}
Each node measures path loss on all links with all other nodes at
each frequency.  A TDMA-based MAC scheme is used in which each node
broadcasts its pairwise measurements during an assigned slot, to
avoid interference.  The data sent by a node in its packet
transmission includes the RSS values recorded during the previous
period, a unique sequence number, its transmit power, and its
battery voltage.


%
%

\showfigure{
\begin{figure}[htbp]
\centerline{
    \psfig{figure=packetStructure_2.eps,width=2.7in}}
\caption{Figure showing the packet structure of each transmitted
message. The message is a packet of length 34 bytes out of which 29
bytes are user programmable. The first 5 bytes if data namely 'Rx
Action', 'Message Type', 'Group ID' and 'Data Length' are not user
programmable.} \label{F:packet_structure}
\end{figure}
}


\subsubsection{Receiver Base}\label{S:receiverbase}
The receiver base is a mica2 node connected to a laptop, loaded with
a special receiver program which synchronizes to the frequency
hopping schedule of the nodes and communicates all the received
packets serially to the laptop for storage and later analysis.

%% file: NCMS_experiment.tex
This section describes the use of the NCMS described in Section
\ref{S:MeasurementSetup} to measure a network deployed in an
ensemble of 15 different environments. These measurements will
enable the statistical analysis and model development in subsequent
sections.

\subsection{Motivation}
Ideally, statistical characterization of the radio channel for
multi-hop networks would proceed as follows: deploy \textit{K}
networks, each with \textit{N} nodes positioned with the identical
geometry in the same type of environment, but each network in a
different place. For example, one might deploy the NCMS in a grid,
in \textit{K} different office buildings.

In reality, its not economical to carry out the measurement campaign
in $K$ different office buildings, mainly because it is difficult to
obtain access to carry out measurement in many different office
areas, and it is difficult to position sensors in exactly the same
geometry without moving obstructions to make space for each node. If
the environment must be altered to measure it, we might as well
randomly alter the entire environment.

In fact, in this campaign, we occupy a single environment and
randomly vary the object locations in that environment. We start by
deploying nodes in an empty classroom in the Merrill Engineering
Building at the University of Utah.  A 4x4 square grid of mica2
nodes is set up with 4 ft (1.22 m) separation between neighboring
sensors. Within this deployment area, different arrangements of
obstructions are randomly generated.

\subsubsection{Random Environment Generation}\label{S:RandomEnvGen}
For reasons of portability, the obstructions used in this campaign were
cardboard boxes of size 61 cm x 41 cm x 61 cm (24 in x 20 in x 25
in). In order to make the boxes significant RF scatterers, we wrap
the cardboard boxes with aluminium foil. Foil-wrapped cardboard
boxes represent metal obstacles which might be present in office
environments.

We generate (in Matlab) random positions for 10 boxes to be placed
in the area of the deployment. The Matlab script is written to
ensure that boxes do not lay on top of any of the 16 sensors (which
are placed on the floor).  Beyond that restriction, the rectangular
boxes may be positioned anywhere in the environment and may be
positioned with either N-S or E-W orientation \ie, with their longer
sides parallel or perpendicular to the X-axis as shown in
Fig.~\ref{F:possibleBoxPlacements}.

\showfigure{
\begin{figure}[htbp]
\centerline{
    \psfig{figure=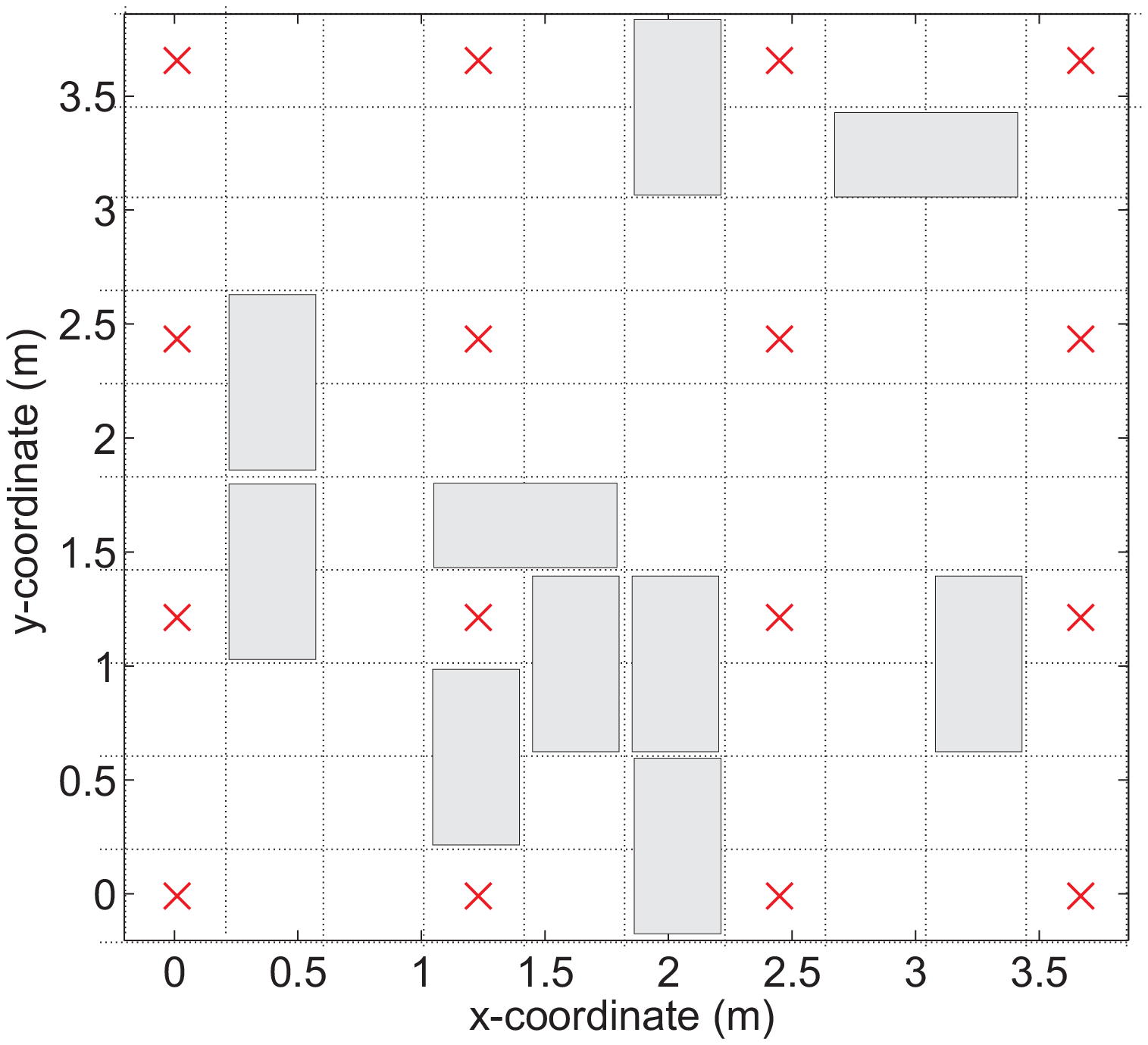,width=2.5in}
    }
\caption{One realization of the random locations of the boxes among
the 16 node locations $\times$. Each box occupies two pixels of this
graph and can be placed either parallel or perpendicular to x-axis.}
\label{F:possibleBoxPlacements}
\end{figure}
}


\subsection{Experiment Procedure}

After random placement of the 10 obstructions, the campaign proceeds
by powering on the 16 nodes and receiving and recording the measured
path loss data in a file on a laptop.  Each node runs the algorithm
described in Section \ref{S:MeasurementSetup}.  After 10 minutes of
run-time, the nodes are turned off. The process continues with the
next measured network by randomly changing the obstruction locations
and repeating the experiment. Fifteen network realizations are
measured in this manner.

%% file: StatisticalAnalysis.tex

This section presents the statistical analysis of the data collected
by campaign described in Section \ref{S:Experiment}. We first
estimate the path loss model parameters of (\ref{E:meanPathLoss})
and (\ref{E:logNormalShadowing}).  Next, we analyze the shadowing
loss correlations which exist on different pairs of links.


\subsection{Analysis of Received Power}\label{S:Analysis_RSS}

We denote the number of the deployment experiment as $m\in
\{1,\ldots,M\}$, where $M$ is the number of deployments (here,
$M=15$).  We denote the set of frequencies measured as
$\mathfrak{F}$.  The received signal power between node $i$ and node
$j$ for experiment $m$ at center frequency $f\in \mathfrak{F}$ is
denoted $P_{i,j}^{(m)}(f)$ and can be written using
(\ref{E:logNormalShadowing}) and (\ref{E:shadowFadingDefn}) as
\begin{equation}\label{E:freqDependentRSS}
    P_{i,j}^{(m)}(f) = P_{T_j} - \Pi_0 - 10n_p\log \frac{d_{i,j}}{\Delta_0} -
    X_{i,j}^{(m)} - Y_{i,j}^{(m)}(f),
\end{equation}
where $Y_{i,j}^{(m)}(f)$ is the non-shadow fading and
$X_{i,j}^{(m)}$ is the shadow fading on link $(i,j)$ during
experiment $m$.  Shadow fading is considered to be constant across
the frequency band, as discussed in Section \ref{S:Introduction}. We
denote the frequency average received power as $P_{i,j}^{(m)}$,
\begin{equation*}
    P_{i,j}^{(m)} \triangleq \frac{1}{|\mathfrak{F}|}\sum_{f\in \mathfrak{F}}
    P_{i,j}^{(m)}(f).
\end{equation*}
From (\ref{E:freqDependentRSS}), we can write $P_{i,j}^{(m)}$ as,
\begin{equation}\label{E:ApproxRSS}
    P_{i,j}^{(m)} = P_{T_j} - \Pi_0 - 10n_p\log \frac{d_{i,j}}{\Delta_0} -
    X_{i,j}^{(m)} - \frac{1}{|\mathfrak{F}|}\sum_{f\in \mathfrak{F}}
    Y_{i,j}^{(m)}(f).
\end{equation}
In other words, (\ref{E:ApproxRSS}) can be written as,
\begin{equation}\label{E:link_receivedSignal_eachExp}
    P_{i,j}^{(m)} = P_{T_j} - \Pi_0 - 10n_p\log \frac{d_{i,j}}{\Delta_0} -
    X_{i,j}^{(m)} - Y_{i,j}^{(m)},
\end{equation}
where $Y_{i,j}^{(m)} = \frac{1}{|\mathfrak{F}|}\sum_{f\in
\mathfrak{F}} Y_{i,j}^{(m)}(f)$.  Because $Y_{i,j}^{(m)}$ is an
average of measurements at many different frequencies, we argue that
it may be well-represented as Gaussian (in dB), regardless of the
underlying frequency-selective fading mechanism (\eg, Rayleigh or
Rician). Since $X_{i,j}^{(m)}$ is also log-normal \cite{coulson}, we expect the sum $Z_{i,j}^{(m)}$ to
also be Gaussian (in dB).

A linear regression of the frequency averaged received signal powers
$\{P_{i,j}^{(m)}\}_{i,j}$ versus known distances $\{ d_{i,j}
\}_{i,j}$ is used to estimate the constants $(P_{T} - \Pi_0)$ and
$n_p$ (\ref{E:link_receivedSignal_eachExp}) for each experiment $m$.
In our experiments, we have used $\Delta_0 = 1$m.  Since all nodes
are set to the same transmit power and have approximately equal
battery voltages, and since we estimate $(P_{T} - \Pi_0)$ in
addition to $n_p$, we are not required to know the exact transmit
power $P_T$ at the current battery voltage of the nodes in the
network during experiment $m$. The linear regression also determines
the variance of $Z_{i,j}^{(m)}$.


\subsection{Analysis of Link Correlations}
\label{S:Analysis_corr}

In this subsection, we describe the computation of the correlation in
fading between pairs of links. This requires computing correlation
in the sample values of $Z_{i,j}^{(m)}$ for different pairs of links
$(i,j)$ as described in Section \ref{S:Analysis_RSS}.

\subsubsection{Similar Geometry Links}
We use the term ``link geometry'' to describe for two links, link
$a$ and link $b$, the relative coordinates of the end points of the
two links. In a grid network, there can be many pairs of links with
the same link geometry (within a rotation). As one example, the link
pair of link $a$ and link $b$ shown in Fig.~\ref{F:Examples}, is
repeated 16 times in the network as shown in
Fig.~\ref{F:SimilarGeoLink}.

\showfigure{
\begin{figure}[htbp]
  \centerline{
  \psfig{figure=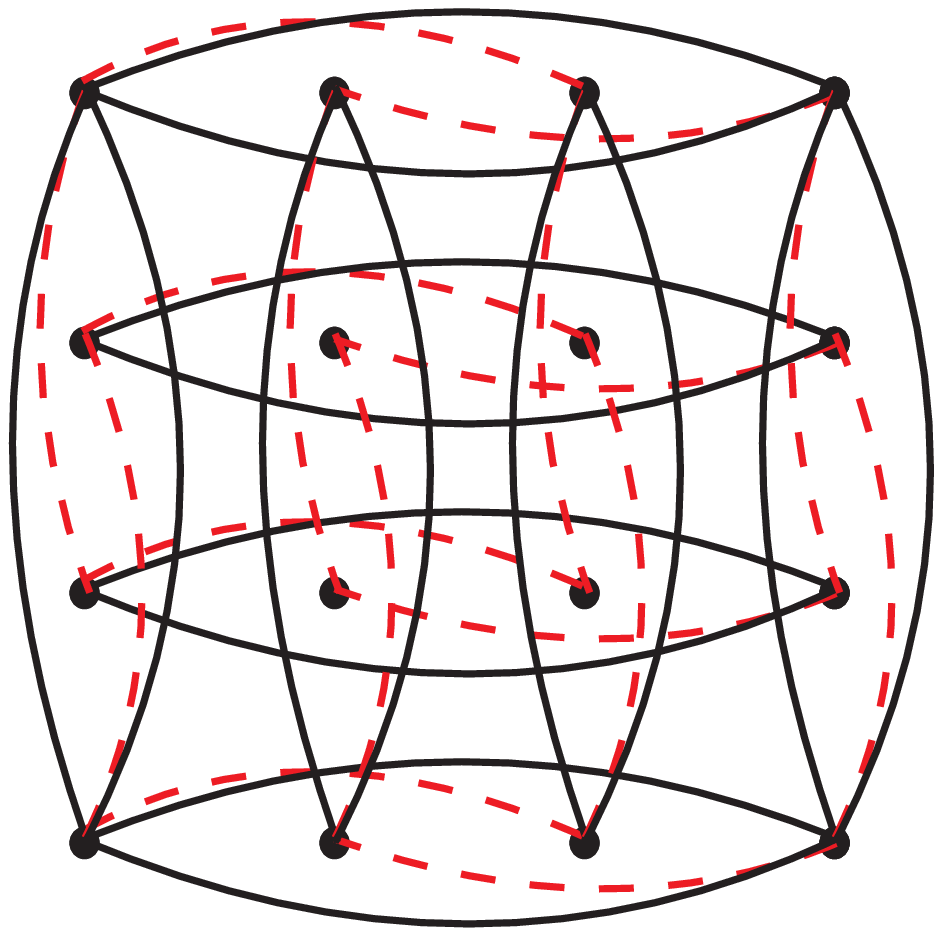,width=1.4in}}
  \caption{Link pairs with identical link geometry in a grid deployment. Each link pair
is shown with one link as a dotted (-~-) line and another link as a
solid lines (--). All link pairs with identical link geometry are
shown.}
  \label{F:SimilarGeoLink}
\end{figure}
}

Let $L$ denote the number of times a particular link geometry is
repeated in the network. We denote the $p^{\mbox{th}}$ link pair as
the two links $(i_p,j_p)$ and $(k_p,l_p)$, where $p\in
\{1,\ldots,L\}$.  Then $Z_{i_p,j_p}^{(m)}$ and $Z_{k_p,l_p}^{(m)}$,
where $m\in \{1,\ldots, M\}$, represent the total fading on the
$p^{\mbox{th}}$ repeated link pair for experiment number $m$. Then
vectors $\mbZ_a^{(m)}$ and $\mbZ_b^{(m)}$ are defined as
\begin{equation}\label{E:Z_ab^mDefn}
    \mbZ_a^{(m)} =
      [Z_{i_1,j_1}^{(m)}, \ldots ,Z_{i_L,j_L}^{(m)}]^T \quad,\quad
    \mbZ_b^{(m)} = [Z_{k_1,l_1}^{(m)}, \ldots
    ,Z_{k_L,l_L}^{(m)}]^T.
\end{equation}
We then define vectors $\mbZ_{a}$ and $\mbZ_{b}$ as
\begin{equation}\label{E:Z_abDefn}
    \mbZ_{a} =
    [\mbZ_{a}^{(1)^T},\ldots,\mbZ_{a}^{(M)^T}]^T \quad,\quad
    \mbZ_{b} =
    [\mbZ_{b}^{(1)^T},\ldots,\mbZ_{b}^{(M)^T}]^T.
\end{equation}
Vectors $\mbZ_{a}$ and $\mbZ_{b}$ are both $LM$x1 sized vectors.
Together they contain all measured total fading values for pairs of
links which share a particular link geometry. The correlation
coefficient of total fading on link $a$ and link $b$,
$\rho_{Z_a,Z_b}$, can be computed by taking vectors $\mbZ_{a}$ and
$\mbZ_{b}$ as sample values of total fading for link $a$ and link
$b$ respectively.

We have computed the correlation coefficient for total fading on
link $a$ and link $b$ for a variety of link geometries.
Table~\ref{T:correlationComparision} shows the results for various
link pair geometries. We also run a hypothesis test to determine if
the measured correlation is statistically significant. This test
compares hypotheses,
\begin{eqnarray*}
    &H_0:& \mbox{$Z_a$ and $Z_b$ have $\rho=0$},\\
    &H_1:& \mbox{$Z_a$ and $Z_b$ have $\rho\neq0$}.
\end{eqnarray*}
We report  $\PR{\mbox{measuring }\rho|H_0}$ using the method
described in \cite[pp. 427-431]{Hines4th2003}, in
Table~\ref{T:correlationComparision}. The proposed correlated link
shadowing model and the Gudmundson model, also mentioned in
Table~\ref{T:correlationComparision}, will be discussed in Section
\ref{S:CorrelationModelDes}.

\begin{table}[p]
\begin{center}
\begin{tabular}{|c|c|l|c|c||c|c|l|c|c|}
  \hline
   & Geometry & \multicolumn{3}{c|}{Correlation $\rho$} & & Geometry & \multicolumn{3}{c|}{Correlation $\rho$}\\
    \hline
            & & Meas-  & Prop. & Gud.  & & & Meas- & Prop. & Gud.\\
            & & ured   & Model & model & & & ured  & Model & model  \\
   \hline
   1 & \psfig{figure=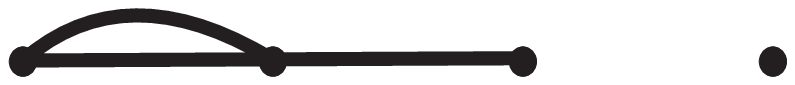,width=0.5in} &0.33*** &0.21 & 0.13 & 15& \psfig{figure=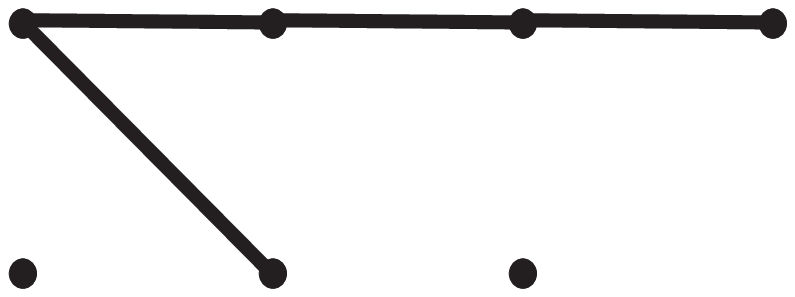,width=0.5in} &-0.04    &0.05& 0.04 \\
   \hline
   2 & \psfig{figure=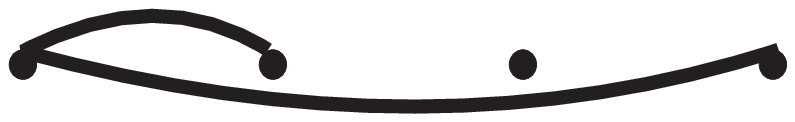,width=0.5in} &0.21*** &0.17 & 0.04 & 16& \psfig{figure=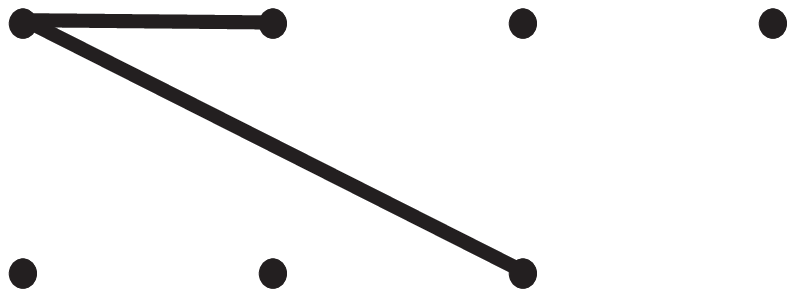,width=0.5in} &0.12*** &0.10& 0.08 \\
   \hline
   3 & \psfig{figure=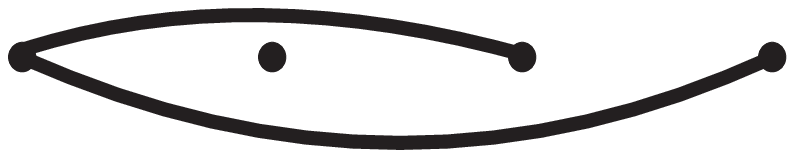,width=0.5in} &0.23*** &0.24 & 0.13 & 17& \psfig{figure=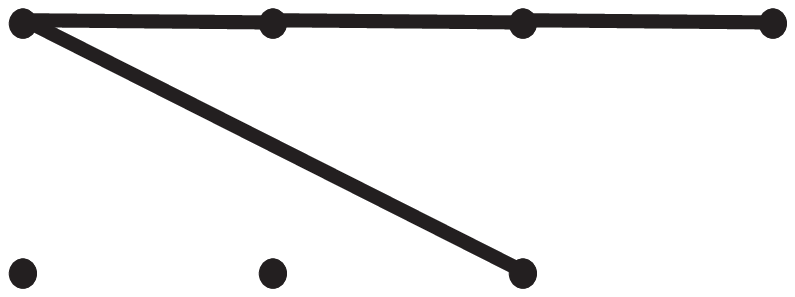,width=0.5in} &0.08*   &0.07& 0.08 \\
   \hline
   4 & \psfig{figure=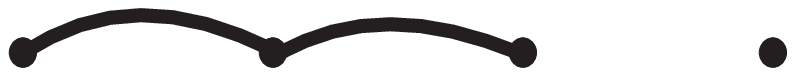,width=0.5in} &0.05     &0.03 & 0.04 & 18& \psfig{figure=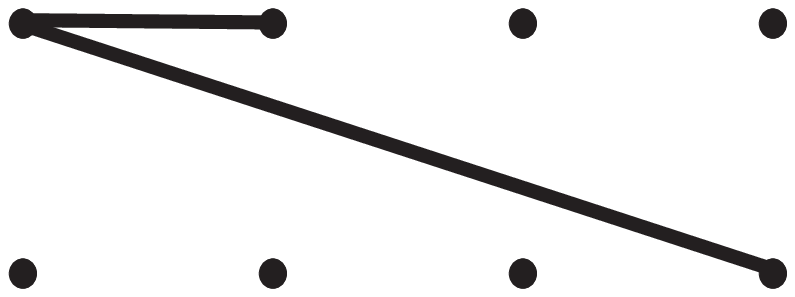,width=0.5in} &0.12*** &0.11& 0.04 \\
   \hline
   5 & \psfig{figure=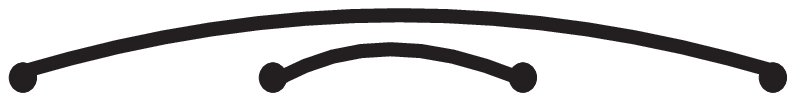,width=0.5in} &0.17*** &0.19 & n/a & 19& \psfig{figure=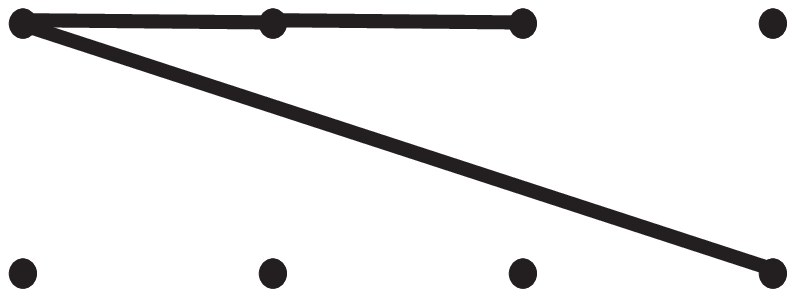,width=0.5in} &0.03     &0.10& 0.08 \\
   \hline
   6 & \psfig{figure=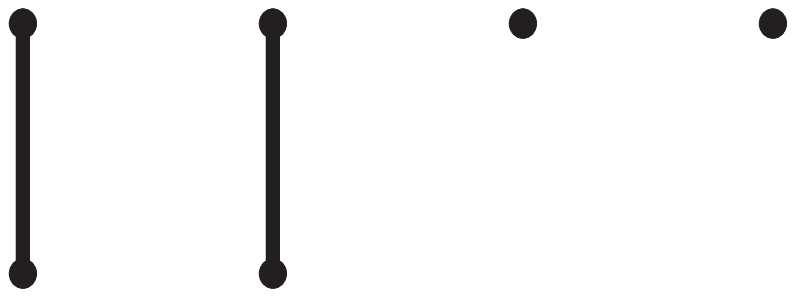,width=0.5in} &-0.05    &0.00 & n/a & 20& \psfig{figure=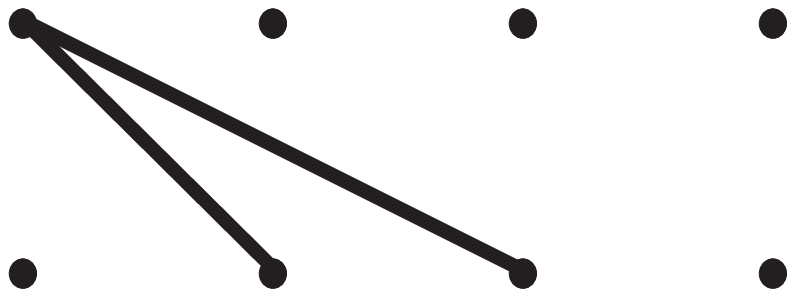,width=0.5in} &0.21*** &0.13& 0.13 \\
   \hline
   7 & \psfig{figure=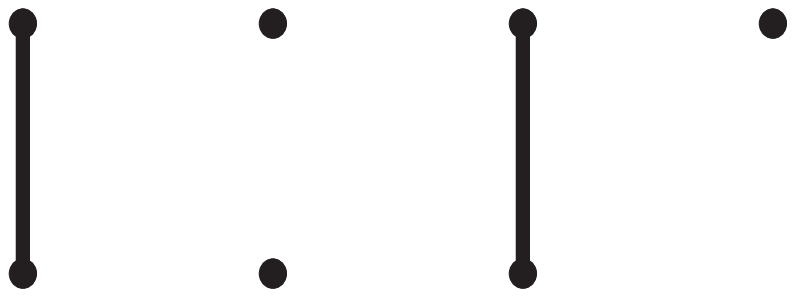,width=0.5in} &-0.01    &0.00 & n/a & 21& \psfig{figure=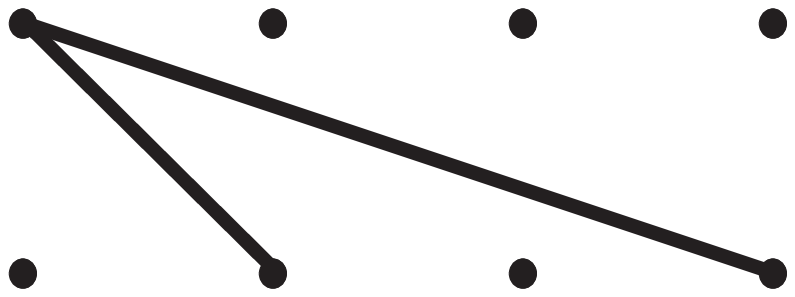,width=0.5in} &-0.02    &0.08& 0.04 \\
   \hline
   8 & \psfig{figure=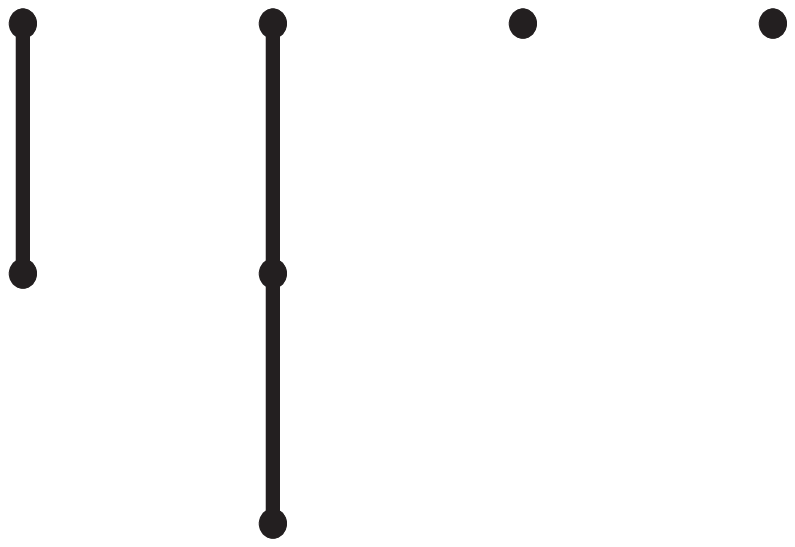,width=0.5in} &-0.10** &0.00 & n/a & 22& \psfig{figure=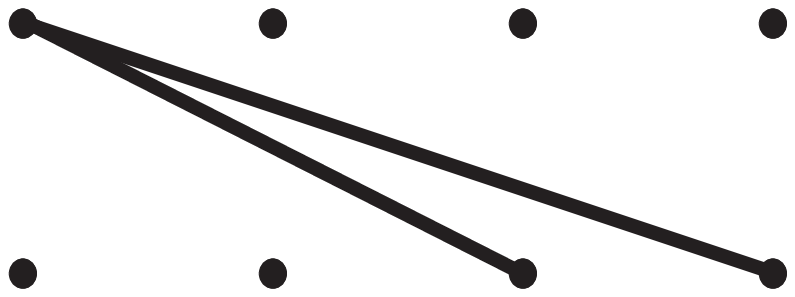,width=0.5in} &0.23*** &0.16& 0.13 \\
   \hline
   9 & \psfig{figure=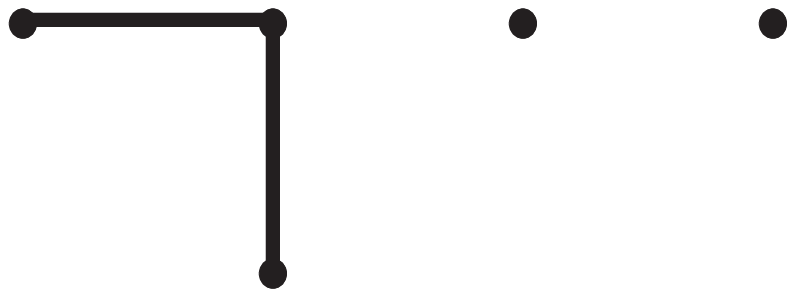,width=0.5in} &-0.03    &0.05 & 0.04& 23& \psfig{figure=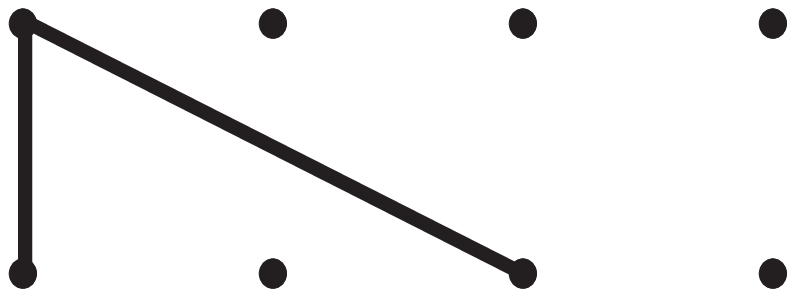,width=0.5in} &0.00     &0.05& 0.04 \\
   \hline
   10&\psfig{figure=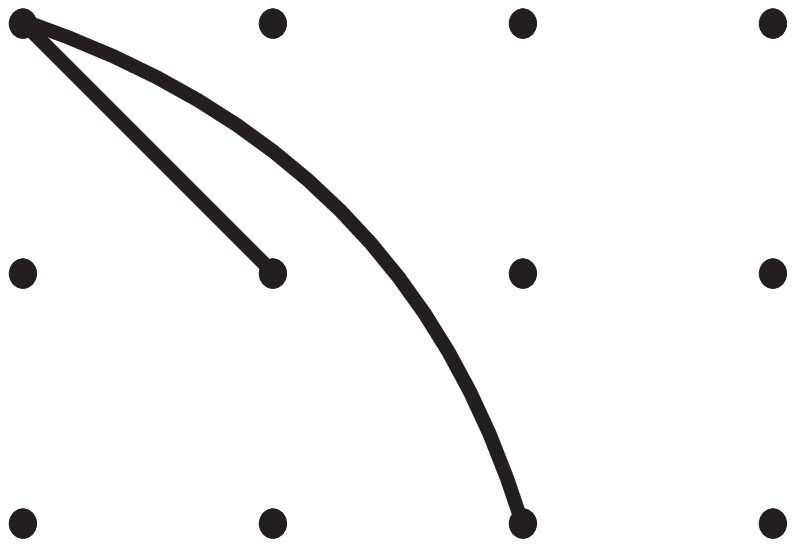,width=0.5in}&0.18*** &0.21 & 0.08& 24& \psfig{figure=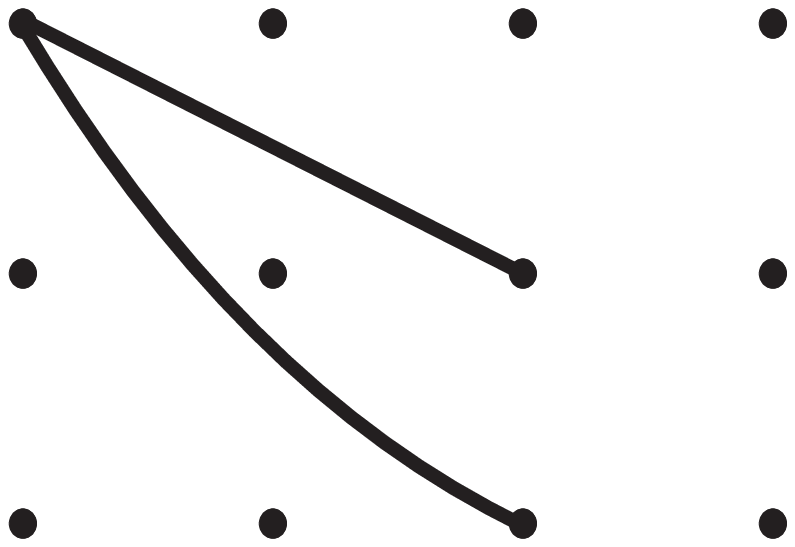,width=0.5in} &0.06     &0.16& 0.08 \\
   \hline
   11& \psfig{figure=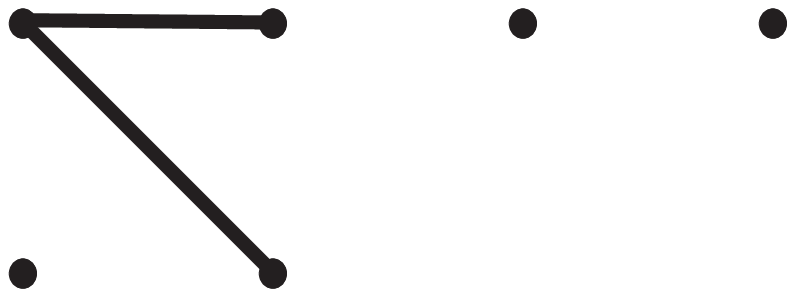,width=0.5in}&0.04*   &0.08 & 0.13& 25& \psfig{figure=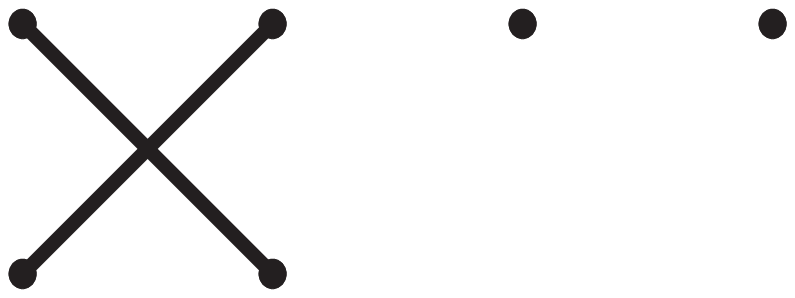,width=0.5in} &0.08**  &0.13 & n/a \\
   \hline
   12& \psfig{figure=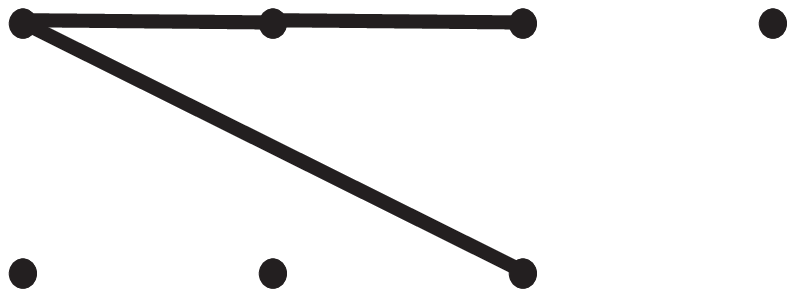,width=0.5in} &0.14***&0.08 & 0.13& 26& \psfig{figure=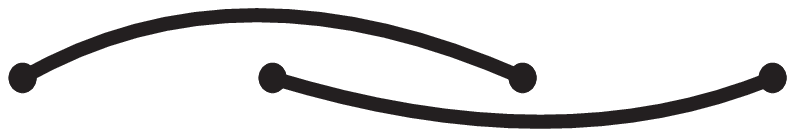,width=0.5in} &0.12     &0.16 & n/a \\
   \hline
   13& \psfig{figure=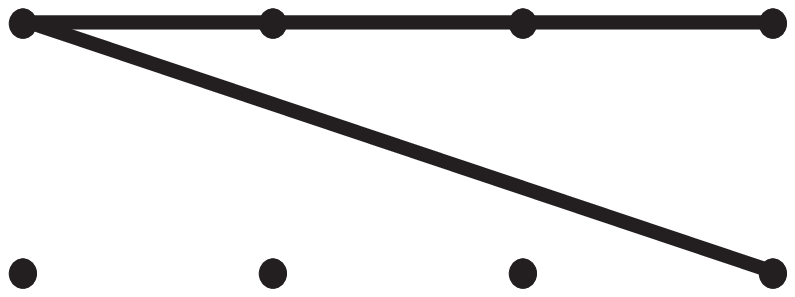,width=0.5in} &0.17***&0.08 & 0.13& 27& \psfig{figure=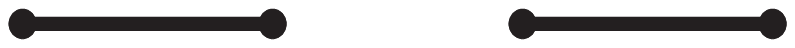,width=0.5in} &0.08     &0.00 & n/a \\
   \hline
   14& \psfig{figure=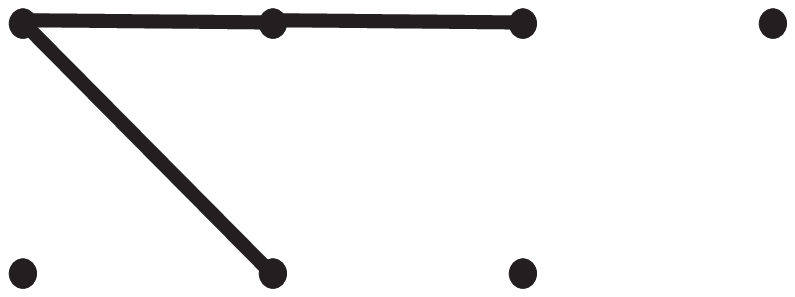,width=0.5in} &0.05    &0.06 & 0.08& 28& \psfig{figure=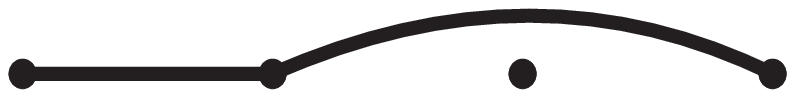,width=0.5in} &0.03     &0.02 & 0.02 \\
   \hline
\end{tabular}
\\
\begin{tabular}{|c c c|}
\hline \multicolumn{3}{|c|}{p- value or $\PR{\text{getting measured
$\rho | H_0$}}$}\\
\hline
    *** $p < 0.005$ & $\quad$ ** $p < 0.01$ & $\quad$ * $p < 0.05$\\
\hline
\end{tabular}
\end{center}
\caption{Link Geometry and Correlation Coefficients (Observed, Proposed Model, and Model of [Gudmundson 1991])}
\label{T:correlationComparision}
\end{table}

\subsection{Discussion}
The results show that, for many link pair geometries, it is
extremely unlikely that the fading losses measured on the pair of
links are independent.  For 15 of the 28 studied link geometries,
there is statistically significant non-zero correlation. Those 15
links are consistently those geometries in which the two links are
proximate, \ie, their lines from transmitter to receiver partially
overlap, or nearly overlap.  The likelihood that the measured
correlation coefficient was measured by chance in the case when
$\rho=0$ is extremely small, \ie, less than 0.5\%, for 11 of the 15
link geometries which showed correlation.

Also note that the correlation coefficients are relatively large in
magnitude.  The highest $\rho$ is 0.33, six link geometries have
$\rho > 0.20$, and eleven link geometries have $\rho > 0.10$. Fading
loss on one link is obviously not purely determined by the losses
experienced on its geographically proximate links; however, the
correlation coefficient indicates that knowing the losses on the
proximate links can give quite a bit of information about the loss
on that one link.

%% file: model_corr_derivation.tex

In this section we present a model to describe the experimentally
observed characteristic of correlated link shadowing.  We start with
the assumption that shadowing loss experienced on the links in a
network is a result of an underlying spatial loss field
$\mbp(\mbx)$, such that shadowing on a link is increased when its
path crosses areas of high loss $\mbp(\mbx)$.  We show how this
assumption results in agreement with existing path loss models when
considering a single link.  We then show how it accurately
represents correlated shadowing losses when jointly considering
links in a multi-hop network.

\subsection{Shadow Fading Model}

In particular, we assume that the underlying spatial loss field
$\mbp(\mbx)$ is an isotropic wide-sense stationary Gaussian random
field with zero mean and exponentially-decaying spatial correlation.
The covariance between $\mbp$ at positions $\mbx_1$ and $\mbx_2$ as
\begin{equation}\label{E:spatial_correlation}
    \E{}{\mbp(\mbx_1)\mbp(\mbx_2)} = R_p(\mbx_1,\mbx_2) = R_p(\|\mbx_2 -
    \mbx_1\|) = \frac{\sigma_{X}^2}{\delta}\exp\left(-\frac{\|\mbx_2 -
    \mbx_1\|}{\delta}\right).
\end{equation}
where $\|\mbx_2 - \mbx_1\|$ is the Euclidian distance between
$\mbx_1$ and $\mbx_2$, $\delta$ is a space constant and $\sigma_{X}$
is the standard deviation of the shadow fading.  The contour plot of
a realization of such a random process is shown in Figure
\ref{F:linkPairInRandomField}.

Many mathematically valid spatial covariance functions are possible
\cite{worsley91}. We justify the use the covariance function in
(\ref{E:spatial_correlation}) because of its basis in a Poisson
spatial random process.  Poisson processes are commonly used for
modeling the distribution of randomly arranged points in space, and
we suppose that attenuating obstructions might be modeled in such a
fashion as well.  Without detailing a specific model for the spatial
extent or value of attenuation of each obstruction, we note that
many Poisson processes (or derivatives of Poisson processes) have
covariance functions with an exponential decay as a function of
distance, as (\ref{E:spatial_correlation}).



\showfigure{
\begin{figure}[htbp]
  \centerline{
  \psfig{figure=linksInRandomField3.eps,width=3in}}
  \caption{A link pair in an underlying spatial loss field}
  \label{F:linkPairInRandomField}
\end{figure}
}

We propose to model the shadowing on all links as functions of the
spatial loss field.  We propose to model the shadowing on link
$(m,n)$, $X_{m,n}$, as
\begin{equation}\label{E:shadowing_link_integral_linkab}
X_{m,n} \triangleq \frac{1}{\|\mbx_n -
    \mbx_m\|^{1/2}}\int_{\mbx_m}^{\mbx_n}
    \mbp(\mbx)d\mbx.
\end{equation}

\paragraph{Single-Link Properties}
This model agrees with two important empirically-observed link
shadowing properties:
\newcounter{Lcount}
\begin{list}{Prop-\Roman{Lcount}}
    {\usecounter{Lcount}
    \setlength{\rightmargin}{\leftmargin}}
    \item The variance of dB shadowing on a link is
      approximately constant with the path length
      \cite{coulson},\cite{hashemi93},\cite{rappaport96}.
    \item Shadow fading losses are Gaussian.
\end{list}
The model in (\ref{E:shadowing_link_integral_linkab}) can be seen to
have Prop-II, since $X_{m,n}$ is a scaled integral of a Gaussian
random process.

The proposed model has Prop-I when $\|\mbx_j - \mbx_i\| >> \delta$.
We show this by considering the variance of $X_a$,
\begin{equation}\label{E:var_shadowfading_linka}
\begin{split}
        \Var{}{X_a} & = E[X_{a}^2]\\
                    & = \frac{1}{\|\mbx_j -\mbx_i\|} \int_{\mbalpha =
                        \mbx_i}^{\mbx_j} \int_{\vbeta =
                        \mbx_i}^{\mbx_j} R_p(\|\vbeta -
                        \mbalpha\|)d\mbalpha^T d\vbeta.
\end{split}
\end{equation}
Using (\ref{E:spatial_correlation}) as the model for spatial
covariance, (\ref{E:var_shadowfading_linka}) is given by
\begin{equation}\label{E:Simplified_var_linka}
    \Var{}{X_a} = \sigma_{X}^2\left[1 + \frac{\delta}{\|\mbx_j - \mbx_i\|} e^{-\|\mbx_j -
                        \mbx_i\|/\delta} - \frac{\delta}{\|\mbx_j -
                        \mbx_i\|}\right].
\end{equation}
When $\|\mbx_j - \mbx_i\| >> \delta$,
\begin{equation}
    \Var{}{X_a} \approx \sigma_X^2.
\end{equation}

\paragraph{Joint Link Properties}

Next, consider two links  $a=(i,j)$ and $b=(k,l)$, as shown in
Fig.~\ref{F:linkPairInRandomField} with shadowing $X_a$ and $X_b$,
respectively. Consider the covariance of $X_a$ and $X_b$,
\begin{equation}\label{E:cov_shadowing_model}
    \Cov{X_{a},X_{b}} =
    \frac{\sigma_X^2}{\delta \|\mbx_i-\mbx_j\|^{1/2}\|\mbx_k-\mbx_l\|^{1/2}}\int_{C_{i,j}} \int_{C_{k,l}}
    e^{-\frac{\|\vbeta - \mbalpha\|}{\delta}} d\mbalpha^T d\vbeta.
\end{equation}
where $C_{m,n}$ is the line between points $\mbx_m$ and $\mbx_n$.
Since $E[X_a] = E[X_b] = 0$, the correlation coefficient between
$X_a$ and $X_b$, $\rho_{X_a,X_b}$, is
\begin{equation}\label{E:correlation_coeff_model}
\begin{aligned}
    \rho_{X_a,X_b} & = \frac{\Cov{X_a,X_b}}{\sqrt{\Var{}{X_a}
    \Var{}{X_b}}}\\
    \rho_{X_a,X_b} & \approx \frac{1}{\delta \|\mbx_i-\mbx_j\|^{1/2}\|\mbx_k-\mbx_l\|^{1/2}}
    \int_{C_{i,j}} \int_{C_{k,l}}e^{-\frac{\|\vbeta - \mbalpha\|}{\delta}} d\mbalpha^T d\vbeta.
\end{aligned}
\end{equation}
The solution to (\ref{E:correlation_coeff_model}) is tedious to
analytically derive. We use numerical integration to compute the
value of $\rho_{X_a,X_b}$, and Matlab calculation code is available
on the authors' web site \cite{code_download}.


\subsection{Total Fading Model}
Since shadowing loss $X_{i,j}$ is only one part of the total fading
loss $Z_{i,j} = X_{i,j} + Y_{i,j}$, we must also consider the model
for non-shadowing losses $Y_{i,j}$.  It is worthwhile to note that
shadow fading and non-shadow fading are caused by different physical
phenomenon, and thus $X_{i,j}$ and $Y_{i,j}$ can be considered as
independent. The variance $\Var{}{Z_{i,j}}$ is thus
\begin{equation}\label{E:Variance_totalFaiding_linka}
    \sigma_{dB}^2 \triangleq \Var{}{Z_{i,j}} = \Var{}{X_{i,j} + Y_{i,j}} = \Var{}{X_{i,j}} +
    \Var{}{Y_{i,j}}.
\end{equation}
Non-shadow fading is predominantly composed of frequency-selective
or small-scale fading, which can be well-approximated to have zero
correlation over distances greater than a few wavelengths.  Since
multi-hop networks typically have sensors spaced more than a few
wavelengths apart, $\{Y_{i,j}\}$ are considered independent in this
paper.

Thus the correlation coefficient between the total fading on links
$a$ and $b$, $Z_a$ and $Z_b$, is
\begin{equation}\label{E:corr_coeff_totalFading}
\begin{split}
    \rho_{Z_a,Z_b} & = \frac{\Cov{Z_a,Z_b}}{\sqrt{\Var{}{Z_a}
                        \Var{}{Z_b}}}\\
    & = \left\{
       \begin{array}{ll}
         1, & \text{if $a = b$} \\
         \frac{\sqrt{\Var{}{X_a}\Var{}{X_b}}}{\sigma_{dB}^2} \rho_{X_a,X_b} \approx \frac{\sigma_X^2}{\sigma_{dB}^2}\rho_{X_a,X_b}, & \text{if $a \neq b$}
       \end{array}
     \right.
\end{split}
\end{equation}
Equation (\ref{E:corr_coeff_totalFading}) indicates a linear
relationship between the correlation coefficient of total fading and
correlation coefficient of shadow fading. The correlation
coefficient, $\rho_{Z_a,Z_b}$, is the measured $\rho$ computed in
Table~\ref{T:correlationComparision}.  The total fading variance
$\sigma_{dB}^2$ was determined by the regression analysis in Section
\ref{S:Analysis_RSS}.

\subsection{Estimation of model parameters from measurements}
Both the space constant $\delta$ and the variance of shadowing
$\sigma_X^2$ must be determined experimentally from the data set.
Specifically, we find the $(\delta, \sigma_X^2)$ pair which best
explains the correlations which exist in the link measurements. In
other words, the goal is to find the value of $(\delta, \sigma_X^2)$
that results in highest agreement between measured and model-based
correlation values.

To accomplish this model fitting, we compute the model correlation,
$\rho_{X_a,X_b}$, for a range of $\delta \in [0.1, 0.4]$ using
(\ref{E:correlation_coeff_model}), for each of the 28 link
geometries considered in Table \ref{T:correlationComparision}.  At a
particular value of $\delta$, we compare the model correlation
$\rho_{X_a,X_b}$ with the measured correlation $\rho_{Z_a,Z_b}$
using linear regression.  This linear regression returns a
correlation coefficient, $\rho_C$, which quantifies how well the
model (using $\delta$) agrees with the measurements. The highest
value of $\rho_C$ uses $\delta^*$, the optimum $\delta$ which
matches the model to the measurements.
Fig.~\ref{F:corr_varyingDelta} plots the correlation $\rho_C$ for
$\delta \in [0.1 ~ 0.4]$. We can observe that the curve attains the
maximum $\rho_C$ at $\delta^* = 0.21$.  The value of $\sigma_X^2$ is
then determined from $\delta^*$ using
(\ref{E:corr_coeff_totalFading}), and we see that
$\sigma_X^2/\sigma_{dB}^2 = 0.29$.

\showfigure{
\begin{figure}[htbp]
  \centerline{
  \psfig{figure=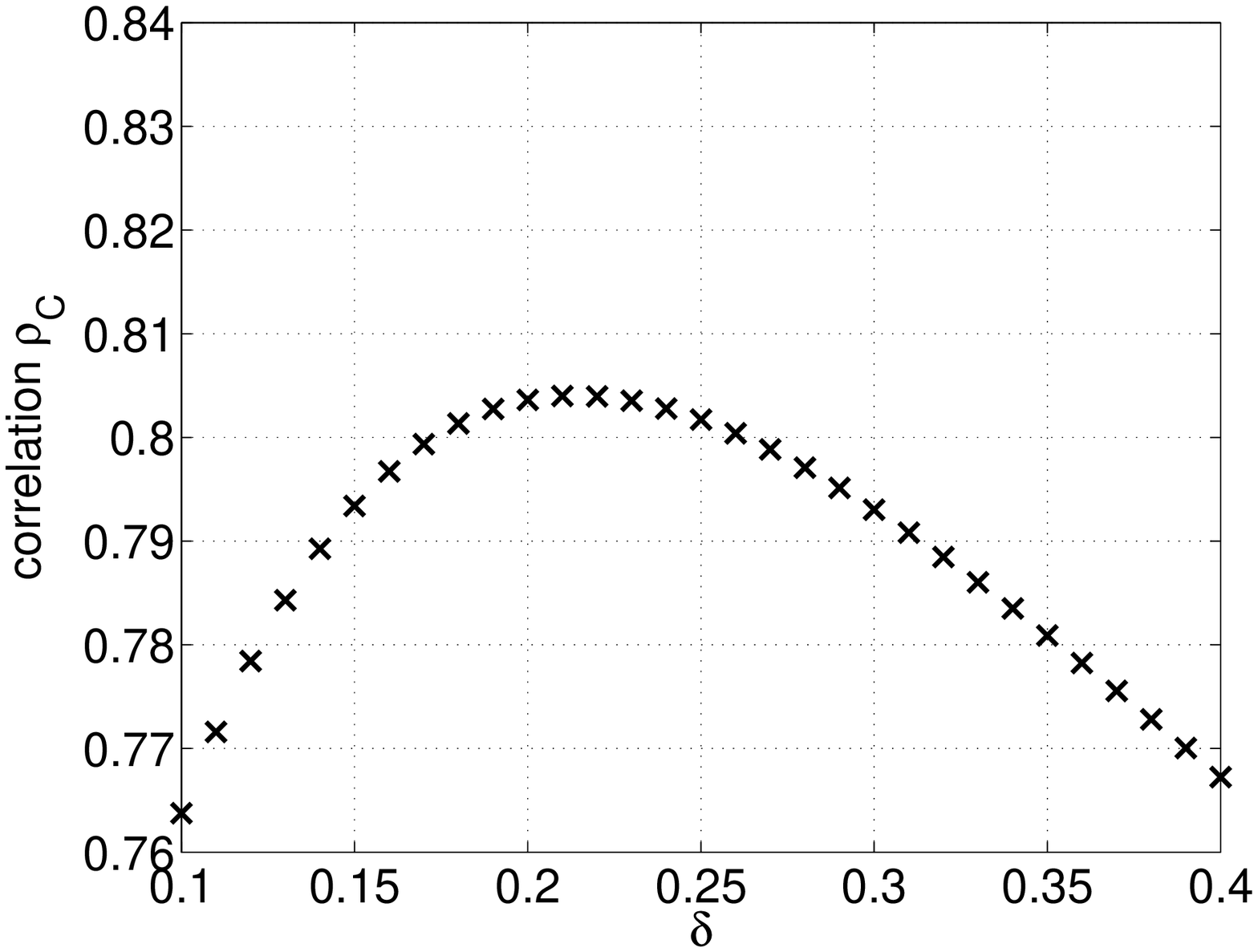,width=2.7in}}
  \caption{Variation of $\rho_C$ with $\delta$. }
  \label{F:corr_varyingDelta}
\end{figure}
}

In summary, we have determined the two parameters of the correlation
model, $(\delta, \sigma_X)$ using the measurement data set.


\subsection{Comparison with Gudmundson Model} In this section, we
compare the proposed model of shadow fading correlation with an
application of an existing model \cite{Gudmundson91}. Gudmundson's
model addresses cellular radio networks where a mobile receiver (low
antenna) communicates with a base station (high antenna). As the
mobile receiver changes position with respect to the base station as
shown in Fig.~\ref{F:mobileBSPosition}, there can be significant
correlation in shadowing on the links with the base station. For a
mobile receiver moving with a velocity $v$, and sampling signals at
every $T$ seconds, the correlation in shadowing $R_X(k)$ is given as
\begin{equation}\label{E:GudmundsonModel}
  R_X(k) = \sigma^{2}_X a^{|k|} \quad \text{where,}\quad a = \epsilon^{v\textsc{T}/\textsc{d}}_\textsc{d}.
\end{equation}
where $D$ is the reference distance, $\epsilon_D$ is the correlation
in shadowing on links when the mobile receiver moves a distance $D$,
and $\sigma^{2}_X$ is the variance of the shadowing on a link.
\showfigure{
\begin{figure}[htbp]
  \centerline{
  \psfig{figure=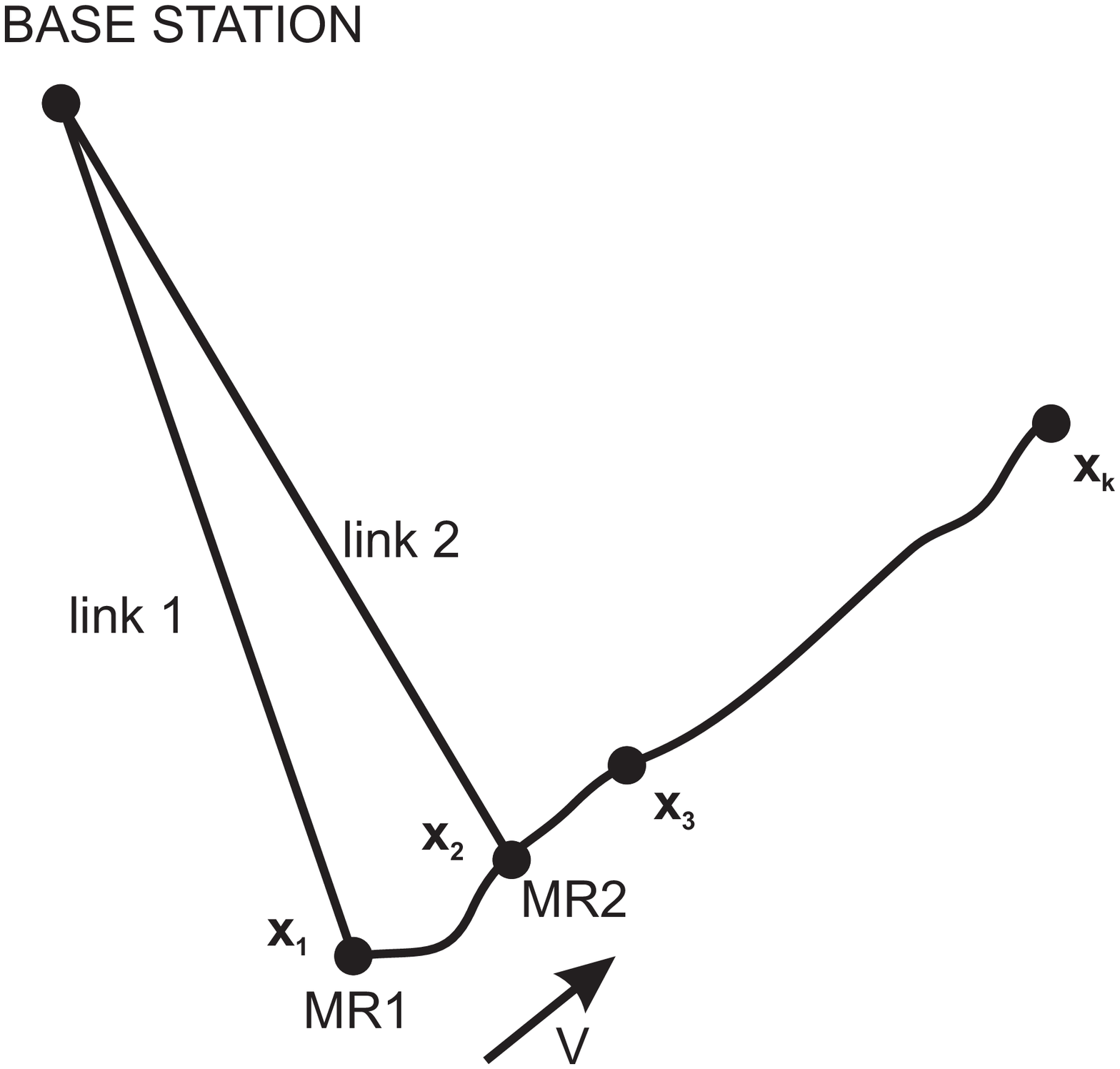,width=2.5in}}
  \caption{Example of the motion of mobile receiver and base station
  position in Gudmundson's model}
  \label{F:mobileBSPosition}
\end{figure}
}

\subsubsection{Application to Multi-hop Networks}

Because the model of (\ref{E:GudmundsonModel}) was not designed for
ad hoc networks, it can only be applied to pairs of links which
share a common node.  This is a major limitation of the Gudmundson
model which requires development of a new shadowing correlation
model for multi-hop networks.  Regardless, we consider here the
application of (\ref{E:GudmundsonModel}) to pairs of links which
share a common node.  The shadowing correlation between the two
links from a common node to two nodes at $\mbx_i$ and $\mbx_j$ can
be written as
\begin{equation}\label{E:modifiedGudmundsonCorr}
  R_X(\mbx_i, \mbx_j) = \sigma^{2}_X \epsilon^{\|\mbx_i -
  \mbx_j\|/\textsc{d}}_\textsc{d}.
\end{equation}
Taking the logarithm of (\ref{E:modifiedGudmundsonCorr}), we get a
linear equation in $\|\mbx_i - \mbx_j\|$,
\begin{equation}\label{E:LinearEq_gudmun}
    \log R_X(\mbx_i, \mbx_j) = \log \sigma^{2}_X + \frac{\|\mbx_i - \mbx_j\|}{D} \log \epsilon_\textsc{d}.
\end{equation}
The constants $\sigma_X^2$ and $\epsilon_\textsc{d}$ can be
determined by running a linear regression between $\log R_X(\mbx_i,
\mbx_j)$ and measured correlation values (in Table
\ref{T:correlationComparision}).

Another limitation of applying the Gudmunson model to multi-hop
networks is that it ignores the location of the common node. For
example, in the two examples in Fig.~\ref{F:limit_Gudmundson}, the
correlation predicted by Gudmundson's model would be identical for
both (a) and (b).  Experimentally, the correlation varies
significantly, from 0.21 in (a)  to 0.05 in (b).   Gudmundson's
model is based on the assumption that the distance between the base
station and mobile station is large compared to distance moved by
the mobile.  This assumption is not generally applicable to
multi-hop networks.

\showfigure{
\begin{figure}[htbp]
\centerline{(a)  \psfig{figure=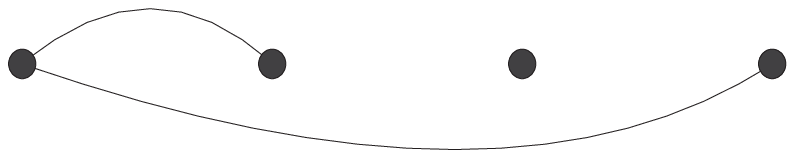,width=1.5in} $\quad$
(b)  \psfig{figure=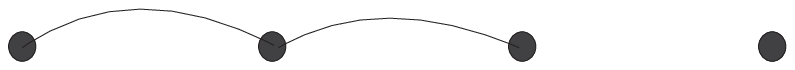,width=1.5in}
}
  \caption{A case of two different types of links, shown by (a)
  and (b). The Gudmundson's model predicts identical correlation for the two
  cases while the proposed model does not. Experimentally, the
  correlations vary significantly from (a) 0.21 to (b) 0.05}
  \label{F:limit_Gudmundson}
\end{figure}
}

Table~\ref{T:comapr_propModelNGudmundsonModel} compares the ability
of the proposed and Gudmundson's model to predict the measured
correlation values.  For the proposed model, we compare the $\rho$
value from the `Prop. Model' column of Table
\ref{T:correlationComparision} with the `Measured' column, for all
28 link geometries tested.  For Gudmunson's model, we compare the
`Gud. model' with the `Measured' $\rho$ for the 21 link geometries
to which the model can be applied. We observe that the measurements
have 80.4\% agreement with the proposed model, compared to 64.4\%
with Gudmunson's model.  Note that while both models are `fit' to
the data, the comparison is valid since both models require fitting
of two parameters ($\sigma_X$ and $\delta$ in proposed model and
$\sigma_X$ and $\epsilon_\textsc{d}$ in the Gudmundson model) to the
data.

\showfigure{
\begin{table}[phtb]
\begin{center}
\begin{tabular}{|l|c|}
  \hline
  & Correlation with Measured Data\\
  \hline
  Proposed Model & 0.804 \\
  \hline
  Gudmundson's Model & 0.644 \\
  \hline
\end{tabular}
\caption{Comparison between the proposed model and Gudmundson's
model}\label{T:comapr_propModelNGudmundsonModel}
\end{center}
\end{table}
}

%% file: application_model.tex

In this section, we study the effects of shadow fading correlation
in two fundamental multi-hop network examples, paths in three and
four node ad-hoc networks. We show by analysis and simulation that
the probability of a path failure can be significantly higher when
links have correlated, as opposed to independent, link shadowing.

To simplify the analysis we assume
\begin{enumerate}
\item Packets are received if and only if the received power is
greater than a threshold $\gamma$, and
\item No packets are lost due to interference.
\end{enumerate}
These assumptions do not limit the results in this section. In fact,
performance in interference is also affected by joint path losses,
and is further impacted by correlated shadowing.

We denote the \textit{normalized received power above the threshold}
for a link $(m,n)$, as $\beta_{m,n}$,
\begin{equation}\label{E:betadefn}
    \beta_{m,n} = \frac{P_{m,n} - \gamma}{\sigma_{dB}}
\end{equation}
where $\gamma$ is the threshold received power and $P_{m,n}$ is
received power given in (\ref{E:logNormalShadowing}). Link $(m,n)$,
by assumption, is connected if and only if $\beta_{m,n}
>0$. An important system parameter is the expected value of $\beta_{m,n}$,
\begin{equation}\label{E:betamean}
    \bar{\beta}_{m,n} \triangleq \E{}{\beta_{m,n}} = \frac{\bar{P}(d_{m,n}) -
    \gamma}{\sigma_{dB}}\\
\end{equation}
where $\bar{P}(d_{m,n})$ is given in (\ref{E:meanPathLoss}).
Intuitively, $\bar{\beta}_{m,n}$ is the number of standard
deviations of link margin we have in link $(m,n)$. If we design the
multi-hop network with higher $\bar{\beta}_{m,n}$, we will have a
higher robustness to the actual fading in the environment of
deployment. For example, one could set the inter-node distance to
ensure that $\bar{\beta}_{m,n} = 2$, and then link $(m,n)$ would
only be disconnected if total fading loss was two standard
deviations more than its mean.

We define two events relating to the connectedness of links,
\begin{eqnarray}\label{E:eventsAnB_Defn}
\begin{aligned}
  \mathcal{A} & = \{\text{Link $(i,k)$ is connected}\}= \{\beta_{i,k} >
  0\}\\
  \mathcal{B} & = \{\text{Link $(i,j)$ and link $(j,k)$ is
  connected}\}= \{\beta_{i,j} > 0\} \cap \{\beta_{j,k} >
  0\}.
\end{aligned}
\end{eqnarray}
Then $\mathcal{A}\cup \mathcal{B}$ is the event that two nodes $i$
and $k$ can communicate, either directly or through an intermediate
node $j$. We call the probability that node $i$ and $k$ cannot
\emph{not} communicate as the \emph{probability of path failure},
\begin{equation}\label{E:linkProb}
    1 - \PR{\mathcal{A}\cup \mathcal{B}} = 1 - [\PR{\mathcal{A}} +
    \PR{\mathcal{B}} - \PR{\mathcal{A}\cap \mathcal{B}}].
\end{equation}


\subsection{A Three Node Multi-Hop Path}
Consider the simple multi-hop path shown in Fig.~\ref{F:simpleLink}(a),
which represents a part of a typical multi-hop network. In this
example, $||\mbx_i-\mbx_j|| = ||\mbx_j-\mbx_k||$. For node $i$ to transmit information to node
$k$, the message packet can take two routes. One is the direct link
$(i,k)$ and the other is a two hop path through a relay node $j$
\ie, through link $(i,j)$ and through link $(j,k)$.  If for our
particular deployment, the link $(i,k)$ fails due to high shadowing,
there is a chance that the message can still arrive via links
$(i,j)$ and $(j,k)$.  This section shows that this `link diversity'
method is not as robust as would be predicted assuming independent
link shadowing.

\showfigure{
\begin{figure}[htbp]
\centerline{
(a)  \psfig{figure=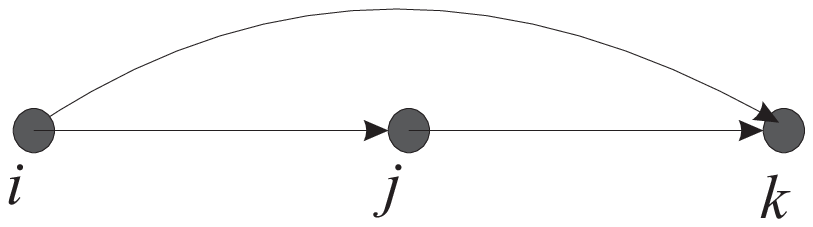,width=1.5in} $\quad$
(b)    \psfig{figure=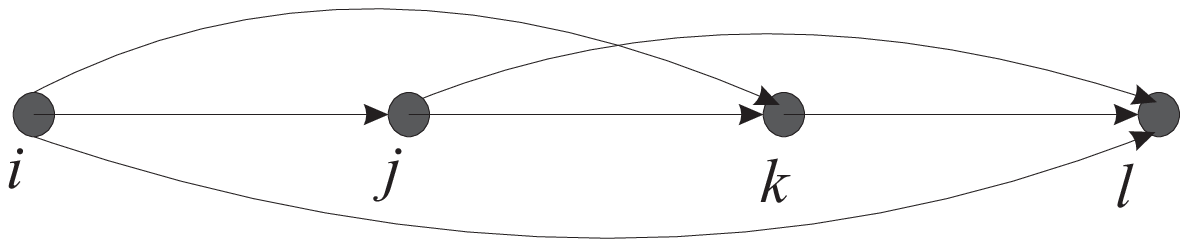,width=2.2in}}
  \caption{Example multi-hop networks of (a) 3 nodes and (b) four nodes.}\label{F:simpleLink}
\end{figure}
}

From (\ref{E:betamean}) and (\ref{E:meanPathLoss}), the relationship
between $\bar{\beta}_{i,j}$, $\bar{\beta}_{j,k}$ and
$\bar{\beta}_{i,k}$ is
\begin{equation}\label{E:Relation_3betaMean}
    \bar{\beta}_{i,j} = \bar{\beta}_{j,k}; \quad \mbox{and} \quad
    \bar{\beta}_{i,k} = \bar{\beta}_{i,j} - \kappa.
\end{equation}
where $\kappa = \frac{10 n_p\log_{10} 2}{\sigma_{dB}}$.


According to the definition (\ref{E:eventsAnB_Defn}), the probability of event $\mathcal{A}$ is,
\begin{equation}\label{E:probA}
\PR{\mathcal{A}} = \PR{\{\beta_{i,k} > 0\}} = \Q{-\bar{\beta}_{i,k}} = 1 -
                            \Q{\bar{\beta}_{i,j} - \kappa}
\end{equation}
where $\Q{\cdot}$ is the complementary CDF of a standard Normal
random variable.

\subsubsection{Case of i.i.d. Shadowing}
Under the assumption that the shadowing across links in a network is i.i.d., the probability of
event $\mathcal{B}$ is
\begin{equation}\label{E:probB_iid}
\PR{\mathcal{B}} = \PR{\{\beta_{i,j} > 0\} \cap \{\beta_{j,k} > 0\}} = (1 - \Q{\bar{\beta}_{i,j}})^2.
\end{equation}
From (\ref{E:probB_iid}) and (\ref{E:linkProb}), the probability of
path failure is
\begin{equation}\label{E:linkfailure_iid}
1 - \PR{\mathcal{A}\cup \mathcal{B}} = \Q{\bar{\beta}_{i,j} - \kappa}\Q{\bar{\beta}_{i,j}}
                                    [2 - \Q{\bar{\beta}_{i,j}}].
\end{equation}


\subsubsection{Case of Correlated Shadowing}
From the correlation values reported in
Table~\ref{T:correlationComparision}, we know that links $(i,j)$ and
$(j,k)$ of Fig.~\ref{F:simpleLink}(a) are nearly uncorrelated. Thus,
the probability for event $\mathcal{B}$ is approximately the same as
in i.i.d. case.  The probability of path failure in this correlated
case is derived in the appendix to be
\begin{equation}\label{E:Prob_AnB_proposed}
    1 - \PR{\mathcal{A}\cap \mathcal{B}} =
    1 - \int_{\beta_{i,k}>0} \left[\Q{\frac{-\mu_1}{\sqrt{1-\rho_{X_{i,j},X_{i,k}}^2}}}\right]^2
    e^{-\frac{(\beta_{i,k} - \bar{\beta}_{i,k})^2}{2}} d\beta_{i,k},
\end{equation}
where,
\begin{equation*}
\nonumber \mu_1 = \bar{\beta}_{i,j} + (\beta_{i,k} - \bar{\beta}_{i,j} +
    \kappa)\rho_{X_{i,j},X_{i,k}}.\\
\end{equation*}

\showfigure{
\begin{figure}[htbp]
\centerline{
  \psfig{figure=plotPercentageIncrProb.eps,width=3in}}
  \caption{Plot showing the simulated and analytical variation of
  percentage increment in the $\PR{\text{path failure}}$ for a
  3 node multi-hop network with normalized received
  power of the shortest link $(i,j)$, $\beta_{i,j}$ }
  \label{F:PercentageIncrProb}
\end{figure}
}


\subsection{A Four Node Multi-Hop Network}

Next, consider the four node link shown in
Fig.~\ref{F:simpleLink}(b). For this linear deployment we assume
$||\mbx_i-\mbx_j||=||\mbx_j-\mbx_k||=||\mbx_k-\mbx_l||$. For node
$i$ to communicate with node $l$, the message packet can be routed
in four ways as shown in Fig.~\ref{F:simpleLink}(b).
An analytical expression for the probability of path failure is
tedious, so instead we simulate the network shown in
Fig.~\ref{F:simpleLink}(b) in both the correlated and i.i.d. link
shadowing.  We take $10^5$ samples of the normalized received powers
under both correlated shadowing and i.i.d shadowing models. We then
determine from the result the probability that there is no path from
node $i$ to node $l$.

 \showfigure{
\begin{figure}[htbp]
\centerline{
  \psfig{figure=plotPercentageIncrProb_4nodes.eps,width=3in}}
  \caption{Plot showing the simulated variation of
  percentage increment in the $\PR{\text{path failure}}$ for a
  4 node multi-hop network with normalized received
  power of the shortest link $(i,j)$, $\beta_{i,j}$}
  \label{F:PercentageIncrProb_4nodes}
\end{figure}
}

\subsection{Discussion}
We compare the probability of path failure between node $i$ and node
$k$ for both the cases of i.i.d.~and correlated link shadowing in
Fig.~\ref{F:PercentageIncrProb_3n4}.  The analysis shows that when a
multi-hop network is designed for $\bar{\beta}_{i,j}=2$, then the
probability of path failure is 120\% greater in correlated shadowing
as compared to i.i.d. shadowing. Increasing the reliability of the
network by designing it for higher $\bar{\beta}_{i,j}$ only
increases the disconnect between the two models. It is only when we
design the network for very unreliable links (\eg,
$\bar{\beta}_{i,j}=0$, for which link $(i,j)$ is connected 1/2 the
time) that the models have a similar result. Clearly, path
connectivity is much more likely under the i.i.d.~model than under
the realistic correlated link shadowing model.

The four-node example shows that as paths become longer, it becomes
increasingly important to consider correlated link shadowing. While
the 3-node network had a 120\% increase in probability of path
failure, the 4-node network showed a 200\% increase in the same
probability.  While Figure \ref{F:PercentageIncrProb_3n4} show the
results up to $\bar{\beta}_{i,j} = 2.5$, higher values correspond to
higher reliability links, and reliable networks will be designed
with even higher link margins. When networks are designed for high
reliability, the effects of ignoring link correlations are dramatic.

%% file: conclusion.tex

A statistical joint path loss model for multi-hop (sensor, ad hoc,
and mesh) networks is presented that relates the shadow fading on
different links in a multi-hop network to the underlying shadowing
field caused by an environment of deployment.  A network channel
measurement system is used to measure a multi-hop network deployed
in an ensemble of environments. The data set is used to demonstrate
and quantify statistically significant shadowing correlations among
different geometries of links.  The measured correlations agree with
the proposed model, and can be applied to a greater variety of links
than possible using an existing correlated shadowing model. Finally,
this paper analyzes path connectivity in simple multi-hop networks
to show the importance of the consideration of shadowing correlation
when designing reliable networks.  The probability of path failure
is underestimated by a factor of two or higher by the current
i.i.d.~shadowing model.

Future work will test other ensembles of deployments, both indoors
and outdoors.  The effects of correlated shadowing will have impact
on higher layer networking protocols and algorithms, and in
interference and multiple-access control, and future work will
quantify this intuition.

%% file: appendix.tex

Here we present the derivation of the probability
$\PR{\mathcal{A}\cap\mathcal{B}}$ in (\ref{E:Prob_AnB_proposed}).
From (\ref{E:betadefn}), we can note that $\beta_{i,j}$,
$\beta_{j,k}$ and $\beta_{i,k}$ are joint Gaussian random variables.
Thus the conditional distributions, $f(\beta_{i,j}|\beta_{i,k} = b)$
and $f(\beta_{j,k}|\beta_{i,k} = b)$, are Gaussian. The links
$(i,j)$ and $(j,k)$ in Fig.~\ref{F:simpleLink} are observed to have
very small or no correlation between them. Thus the joint
distribution, $f(\beta_{i,j},\beta_{j,k}|\beta_{i,k} = b)$, can be
approximated as:
\begin{equation}
f(\beta_{i,j},\beta_{j,k}|\beta_{i,k} = b) \thickapprox
f(\beta_{i,j}|\beta_{i,k} = b)f(\beta_{j,k}|\beta_{i,k} = b).
\end{equation}
The joint distribution of $\beta_{i,j}$, $\beta_{j,k}$ and
$\beta_{i,k}$ is:
\begin{equation}
f(\beta_{i,j},\beta_{j,k},\beta_{i,k}) = f(\beta_{i,j}|\beta_{i,k} = b) f(\beta_{j,k}|\beta_{i,k} = b) f(\beta_{i,k}).
\end{equation}
The probability $\PR{\mathcal{A}\cap\mathcal{B}}$ can be written in
terms of joint distributions as:
\begin{eqnarray}\label{E:Appendix_ProbAnB}
\nonumber \PR{\mathcal{A}\cap\mathcal{B}} &=& \PR{\{\beta_{i,j} > 0\} \cap \{\beta_{j,k} > 0\} \cap \{\beta_{i,k} > 0\}}\\
\nonumber &=& \int_{\{\beta_{i,j} > 0\}} \int_{\{\beta_{j,k} > 0\}}
    \int_{\{\beta_{i,k} > 0\}} f(\beta_{i,j}|\beta_{i,k} = b) f(\beta_{j,k}|\beta_{i,k} = b)
    f(\beta_{i,k}) d\beta_{i,j} d\beta_{j,k} d\beta_{i,k}\\
&=& \int_{b > 0}
    \left[\Q{-\mu_1/\sqrt{1 - \rho_{X_{i,j},X_{i,k}}^2}}\right]^2
    e^{-\frac{(b - \bar{\beta}_{i,k})^2}{2}}
    db.
\end{eqnarray}
where,
\begin{equation*}
\mu_1 \triangleq \mbox{E}[\{\beta_{i,j}|\beta_{i,k} = b\}] =
                        \bar{\beta}_{i,j} + (b-\bar{\beta}_{i,j} +
                        \kappa)\rho_{X_{i,j},X_{i,k}}.
\end{equation*}
The square in the RHS of (\ref{E:Appendix_ProbAnB}) comes from the fact that for the link geometry considered, $\rho_{X_{j,k},X_{i,k}} = \rho_{X_{i,j},X_{i,k}}$.